\documentclass[a4paper,USenglish,cleveref, autoref, thm-restate]{lipics-v2021}

\hideLIPIcs  

\usepackage{comment}
\usepackage{cite}
\usepackage{xurl}
\usepackage{booktabs} 
\usepackage{float}
\usepackage{caption}
\usepackage{makecell}
\usepackage{cleveref}
\crefname{appendix}{appendix}{appendices}
\Crefname{appendix}{Appendix}{Appendices}

\usepackage{algorithm2e}
\usepackage{algorithmic}

\title{Just-in-Time Resale in an Ahead-of-Time Auction for Faster
  Execution}

\author{Burak {\"O}z}%
  {Flashbots, Germany}{burak@flashbots.net}{https://orcid.org/0009-0003-7508-7112}{}
  
\author{Akaki Mamageishvili}%
  {Offchain Labs, Switzerland}%
  {kakia89@gmail.com}{https://orcid.org/0000-0003-2179-7867}{}

\author{Christoph Schlegel}{Flashbots, Switzerland}{christoph@flashbots.net}{https://orcid.org/0000-0001-6727-5031}{}

\author{Ali Taslimi}{Entropy Advisors, UK}{ali@entropyadvisors.com}{}{}

\authorrunning{Burak {\"O}z et al.} 

\Copyright{Burak {\"O}z, Akaki Mamageishvili, Christoph Schlegel, Ali Taslimi} 



\nolinenumbers 

\EventEditors{Aggelos Kiayias and Maria Kyropoulou}
\EventNoEds{2}
\EventLongTitle{8th Conference on Advances in Financial Technologies (AFT 2026)}
\EventShortTitle{AFT 2026}
\EventAcronym{AFT}
\EventYear{2026}
\EventDate{October 6--9, 2026}
\EventLocation{London, UK}
\EventLogo{}
\SeriesVolume{395}
\ArticleNo{36}

\begin{document}

\maketitle

\begin{abstract}

We study Arbitrum's Timeboost auction, an ahead-of-time mechanism that sells a 200\,ms ordering advantage in an otherwise first-come, first-served transaction ordering policy. The market naturally divides into two phases: a \emph{competition phase}, in which the dominant searchers compete directly in the primary auction, and a \emph{coordination phase}, in which they source fast-lane access through Kairos, a Just-in-Time resale intermediary. We use auction bids, time-boosted transactions, and on-chain payment traces to study how well ahead-of-time bids predict realized CEX--DEX arbitrage profits and how the emergence of resale changes surplus allocation.

We find that ahead-of-time bids are noisy predictors of short-horizon arbitrage profits. During the competition phase, bid--profit correlations are statistically significant but economically modest at the round level, while correlations increase when profits are aggregated over longer horizons. This suggests that bidders can identify favorable market conditions but face substantial uncertainty about the realized value of any individual one-minute fast-lane interval. After the Kairos transition, competition in the primary auction weakens sharply: in our focused transition-analysis window, paid bids fall from 62.7\% of the top bid before the transition to 14.8\% thereafter, while total searcher profits remain broadly similar. The resulting dynamics are most consistent with coordination through a common intermediary rather than direct competition in the primary auction. More broadly, our findings suggest that ahead-of-time allocation mechanisms can be vulnerable to secondary-market intermediation when competition among dominant participants is weak.
\end{abstract}

\section{Introduction}
Many blockchains use auctions as part of their transaction ordering mechanism. In Ethereum Layer-1 (L1), {\it Proposer Builder Separation} (PBS) implements a builder auction in which specialized parties called {\it block builders} bid for the right to determine the content of the next proposed block. In most Layer-2 (L2) rollups, transactions are sorted in decreasing order of priority fees per gas. This mechanism is commonly called {\it Priority Ordering} or {\it Priority Gas Auction}, since it amounts to an auction for earlier execution. Despite significant differences---in Ethereum, builders bid for the right to order an entire block while in rollups the auction concerns relative ordering of individual transactions, and Ethereum has a public mempool while rollups do not---these formats share a common feature: they are \emph{Just-in-Time} (JIT), meaning participants can submit bids until immediately before block content is finalized.

An alternative design, \emph{ahead-of-time} (AOT) auctions, introduces a substantial time gap between the auction and the determination of transaction order. In the Ethereum context, such mechanisms have been discussed as slot auctions: a bidder wins the right to propose the next block at auction time but commits to no specific block content until later (typically 12 seconds later)~\cite{slot_auction}. In the rollup ecosystem, Arbitrum's Timeboost~\cite{arbitrum_timeboost_gentle_intro} is the first major deployed instance of a transaction ordering policy with an AOT component.

Prior work on Ethereum slot auctions has argued~\cite{pai2024,trusted} that selling the right to propose blocks in advance may lead to secondary markets for proposal rights that benefit certain searchers. It has also been argued that predicting block value far in advance---as AOT bidders must---is inherently difficult and may lead to inefficient value capture, which can itself be the catalyst for such secondary markets to emerge. As the only major deployed AOT arbitrage auction for transaction ordering, Arbitrum's Timeboost provides a natural empirical case study for evaluating these claims.

\begin{figure}[t]
    \centering
    \includegraphics[width=\linewidth]{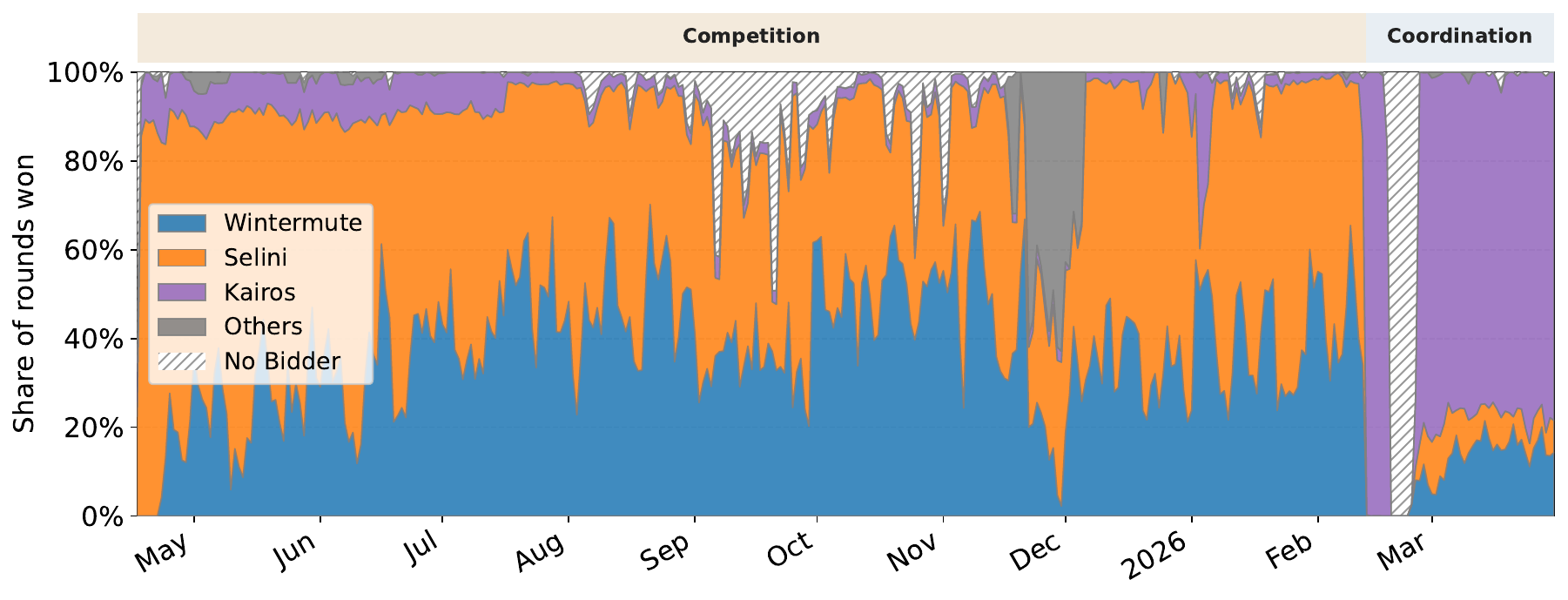}
    \caption{Daily shares of Timeboost auction rounds won by each participant (April 2025–March 2026). The header strip delineates the \emph{competition phase,} in which Wintermute and Selini win the majority of rounds, from the \emph{coordination phase}, in which Kairos takes over as the dominant winner. Hatched regions indicate rounds with no active bidders.}\label{fig:auction_winners}
\end{figure}

Since its launch in April 2025, the Timeboost auction has been dominated by two searchers, Wintermute and Selini Capital, who specialize in non-atomic arbitrage and have won the vast majority of auction rounds (see~\Cref{fig:auction_winners}). Kairos~\cite{kairos_timeboost_docs}, a service that participates in the primary auction and resells fast-lane access at the transaction level JIT, attracted little traction for most of Timeboost's existence. This changed sharply in mid-February 2026, when Wintermute and Selini began sourcing fast-lane access through Kairos' resale market and reduced or ceased primary-auction bidding, after which Kairos began winning the majority of auction rounds.

More precisely, we observe the following sequence of events. On February 12, Wintermute and Selini stopped participating in the primary auction or significantly reduced their bids, leaving Kairos winning nearly all auction rounds at prices close to the reserve. In response, Arbitrum substantially increased the reserve price on February 18, after which no participant was willing to pay it for some period. Bidding later resumed in a small fraction of rounds before the market settled into a new stable regime when Arbitrum lowered the reserve price on February 25. In this regime, Kairos wins approximately 78\% of auction rounds, with occasional but limited participation from Wintermute and Selini.
 
This naturally divides our study into two phases: a \emph{competition phase} prior to February 12, which lets us assess AOT auction effectiveness in the absence of significant JIT resale, and a \emph{coordination phase} thereafter, in which dominant searchers source fast-lane access through a common JIT intermediary rather than competing directly in the primary auction. We use \emph{coordination} in this descriptive sense and do not mean to imply that we observe an explicit agreement among the participants.

We find that, during the competition phase, bids are only weakly correlated with estimated markout arbitrage profit and loss (PnL), consistent with the difficulty of predicting short-horizon value in an AOT setting. We observe that the two dominant searchers route through Kairos's JIT resale market, allowing them to obtain fast-lane access without directly committing to price-uncertain forward bids. Because the associated off-chain payments are unobserved, however, we cannot determine whether this shift increased their net payoffs. In our focused transition-analysis window, paid bids in the primary auction fall to 14.8\% of the top bid after the transition, down from 62.7\% in the February 1--12 pre-transition segment, while estimated arbitrage PnL remain broadly similar across regimes. This indicates that the primary auction captures a smaller share of the value reflected in submitted bids, although the final allocation of surplus between Kairos and the searchers remains unobserved. The observed dynamics are consistent with a shift from direct competition in the primary auction toward routing through a common intermediary. More broadly, our findings suggest that AOT allocation mechanisms may be particularly susceptible to secondary-market intermediation when competition among dominant participants is weak.

\subsection{Related Work}
In Ethereum, the next block proposer sorts transactions unilaterally. This allows the proposer to extract value from the power to order transactions, commonly referred to as Maximum Extractable Value (MEV) ~\cite{MEV}.
Transaction ordering through auctions is theoretically analyzed in~\cite{schlegel2024transaction}. An early version of Timeboost without an AOT auction was proposed in \cite{mamageishvili_et_al:LIPIcs.AFT.2023.23}. Spam levels before and after Timeboost are studied in~\cite{timeboost_spam}, which finds that spam decreased after its introduction. The time advantage of centralized exchange (CEX) to decentralized exchange (DEX) arbitrage under a First-Come-First-Serve (FCFS) ordering policy is studied in~\cite{mev_capture}, with comparisons to PGA auctions and pure FCFS. Latency advantage in a PBS setting are examined in~\cite{latency_advantage,timing_games}. Possible downsides of AOT auctions are identified in~\cite{pai2024} for settings where a secondary market allows JIT resale of AOT-acquired rights.

For methodology, we follow~\cite{cex_dex} in using markouts at CEX mid-prices to estimate CEX--DEX arbitrage profits. Theoretical predictions of such profits are developed in~\cite{milionis2022automated}. Our theoretical background for auctions with common-value component draws on the classical work of~\cite{auction_theory}.

\subsection{Background}

\noindent\textbf{Timeboost.}
Timeboost has been deployed on the Arbitrum rollup since April 17, 2025. Under this mechanism, the protocol sells fast-lane access in a permissionless second-price auction with a reserve price. The fast-lane integrates with Arbitrum's FCFS transaction ordering policy and gives the auction winner a 200\,ms advantage for one minute. The auction for the next minute begins at second 0 of each round and lasts until second 45~\footnote{Each round corresponds to a minute shifted by 9 seconds relative to the wall-clock minute.}. The remaining 15 seconds are used for on-chain auction settlement. The rule is enforced by the sequencer, which executes regular transactions with a 200\,ms delay, while fast-lane transactions from the Timeboost winner are executed immediately upon arrival. In effect, the sequencer merges the two FCFS queues: one for time-boosted transactions and one for regular transactions. Empirically, most fast-lane activity and extracted value can be attributed to CEX--DEX arbitrage~\cite{essias2025express}. In particular, Wintermute exclusively conducts such arbitrage, while Selini occasionally uses fast-lane for other forms of value extraction methods.

\noindent\textbf{Kairos.}
After winning a Timeboost round, Kairos organizes fast-lane ``sub auctions'' with bidding phases of approximately 100\,ms. Transactions within each 100\,ms batch are ordered by profitability, and the batch is submitted to the Arbitrum sequencer via the fast-lane, thereby giving all transactions in the batch a time advantage of at most 100\,ms over regular transactions. The exact advantage depends on the latency between Kairos's servers and the sequencer. As long as this latency remains below 100\,ms, a competing transaction in the regular FCFS queue cannot be executed before the Kairos batch.

Payments to Kairos are typically made on-chain as explicit ETH transfers embedded in time-boosted transactions and sent to a designated contract address\footnote{Kairos payment address: \texttt{0x60E6a31591392f926e627ED871e670C3e81f1AB8}}. An important exception is Kairos's parallel lane~\cite{kairos_parallel_lanes}, which is designed for CEX--DEX arbitrage searchers and operates via a subscription model with off-chain payments. This design bypasses the simulation step in which searchers' transactions are continuously re-submitted under FCFS, thereby reducing latency and lowering the risk of losing priority to regular transactions. Since these transactions constitute a substantial share of all time-boosted transactions in our dataset, the absence of observable on-chain payments prevents us from tracking transfers to Kairos by the two main searchers, as we discuss further below.


\section{Conceptual Framework and Empirical Predictions}
We develop a conceptual framework to organize and guide our empirical
analysis. We first use a standard interdependent-values auction model to discuss how common- and bidder-specific value components affect AOT bidding. We then relate expected arbitrage value to short-run price volatility. Finally, we identify several mechanisms through which a JIT resale market such as Kairos may emerge and derive their distinct empirical implications. The framework is intended to provide qualitative benchmarks rather than a complete structural model of the Timeboost market.


\subsection{Common-Value Second-Price Auctions}
The Timeboost auction has a clear common-value component: the market-wide arbitrage opportunities arising during a Timeboost interval are, in principle, available to all bidders. At the same time, valuations may contain bidder-specific components. Bidders may face private inventory constraints, hold accumulated positions that they want to unwind, or incur heterogeneous trading costs on external markets. For bidders that are also market makers on external exchanges, the trades and inventory accumulated through their market-making activity may further affect their value for winning the auction.
 
A standard framework for auctions with common- and private-value components is the interdependent-values model of~\cite{auction_theory}. Bidders $i = 1, \ldots, n$ receive private signals $X_i$ that are i.i.d., and the value of winning the auction for bidder $i$ is a function of all signals $v_i(X_1,\ldots,X_n)$, which is assumed to be symmetric in opponents' signals, strictly increasing in own signal, and nondecreasing in the others'. Let $Y_i:=\max_{j\neq i}X_j$ denote the highest signal among bidder $i$'s competitors. Under risk neutrality, the symmetric equilibrium bidding strategy in a second-price auction is
\[
b(x)
\;=\;
\mathbb{E}\!\left[\, v_i(X_i, X_{-i}) \;\middle|\; X_i = x,\; Y_i = x \,\right],
\]
the expected value of winning conditional on the pivotal event in which the bidder's signal is exactly tied with that of the strongest competitor.

In contrast to a pure private-value auction, bidders should not simply bid their expected value given their own signal, $\mathbb{E}[v_i(X_i, X_{-i}) \mid X_i = x]$: because the value of winning depends on rivals' signals, winning itself is informative---it reveals that all competing signals were lower---and a bidder who ignores this would systematically overpay. Equilibrium bidding corrects for this winner's curse by evaluating the expected value in the pivotal event where the bidder is exactly tied with the strongest rival.

This framework motivates an empirical comparison between bids and realized arbitrage value. In equilibrium, the price paid equals $b$ evaluated at the second-highest signal, while the top bid reflects the highest signal. If the paid bid is more closely associated with our measure of realized common arbitrage value than the top bid, this is consistent with the common-value component being an important driver of bidder valuations, since the paid bid reflects the signal of the strongest losing bidder, which is informative about realized value only to the extent that values are interdependent. Conversely, if valuations are dominated by private idiosyncratic components, we would expect bids to be less closely associated with realized common arbitrage value. We emphasize that this comparison is suggestive rather than a sharp test: the theory does not imply in general that the second-highest signal must predict realized common value better than the highest, as the relative predictive power of the two order statistics depends on the joint distribution of signals and values.

\subsection{Arbitrage Value and Price Volatility}
The theory of arbitrage in automated market makers (AMM)~\cite{milionis2022automated} can inform how much arbitrage value searchers can expect to extract over a fast-lane interval. While the theory applies to the case where bidders do not face uncertainty about the amount of arbitrage their individual trades will extract~\footnote{Under the Timeboost policy, fast-lane transactions do not know the exact state of the AMMs, as unobserved slow-lane transaction might change it compared to the state that the arbitrageur last observed.}, quantitatively, we should expect similar results. The theory predicts that arbitrage value is proportional to price variance, i.e., expected arbitrage gains over a window of length $T$ scale with price variance over the same window.

Cryptocurrency price volatility is typically non-stationary and is best modeled by a mean-reverting stochastic volatility process~\cite{volatility_mean_reverting}. Therefore, this framework motivates the following empirical expectations. First, if price variance is linear in time, then arbitrage gains are also linear in the length of the Timeboost interval. Second, longer run averages of arbitrage gains should be easier to predict than realized gains in any short interval. This makes value prediction particularly difficult in an AOT auction, where bidders must price access before observing the exact opportunities that will arise.

\subsection{Resale Market}\label{sec:explain}
We identify three main mechanisms through which a JIT resale market such as Kairos could be successful.

\begin{enumerate}
\item {\bf Unbundling}: The Timeboost auction wholesales a 200\,ms time advantage for an entire minute. Reselling this advantage over shorter time intervals or on a per transaction basis could, in principle, enhance efficiency by allowing different searchers to obtain time-boosts within the same minute. Ideally, this would allocate the time advantage to the searcher that can extract the most value at a given moment, rather than to a single party that extracts the most value on average over the full minute. Similar dynamics arise in L1 block-building auctions~\cite{flashbots_mevboost}, where builders resell blockspace to many different searchers. In this case, additional surplus from resale would come from improved market efficiency.

\item {\bf Non-Competitive Behavior}: In the Timeboost setting, the presence of only two dominant searchers makes the market vulnerable to tacit coordination or collusive behavior. Kairos may effectively, though not necessarily deliberately, operate as a hub in a hub-and-spoke coordination structure: the two major searchers may abstain from bidding, or bid less aggressively than they otherwise would, allowing Kairos to acquire the fast-lane at a lower price than under competitive bidding. In this case, the combined surplus of the two searchers and the reseller could exceed that in the competitive benchmark, not because more value is created, but because less revenue is paid to the seller in the primary auction, namely Arbitrum.

\item {\bf Information}: Bidding AOT for the right to use the fast-lane over an entire minute gives bidders less information about the value they will ultimately be able to extract than bidding JIT for an immediately submitted transaction. This matters when some opportunities are searcher-specific, e.g., because a searcher identifies or can exploit opportunities that others cannot. In such cases, a near-JIT auction can generate higher expected revenue\footnote{The mathematical intuition is simple: in the extreme case of a pure private-value auction, the mean of i.i.d.\ random variables is smaller than their expected maximum, so allocating closer to the point of realization can increase expected value extraction.}. A related but separate issue arises if the reseller is also a searcher themselves, for which there is no indication that Kairos qualifies\footnote{\cite{pai2024} focus particularly on this case and argue that it can be a driver of market centralization in the searcher/builder market.}. In that case, the reseller could strategically choose to resell the time-boost only after receiving an unfavorable signal about extractable value, thereby capturing even more value from operating the resale market. In this setting, additional surplus from resale would again arise through improved allocation efficiency.
\end{enumerate}

These mechanisms imply different empirical patterns. The unbundling mechanism predicts a broader and more differentiated set of active searchers under Kairos, together with higher total value extraction through time-boosted transactions. The weakened-competition mechanism predicts that the same dominant searchers remain active and total value extraction stays broadly similar, while revenue in the primary Timeboost auction declines. It may also imply that the dominant searchers stop bidding competitively against each other in the primary auction. The information mechanism predicts higher value extraction from time-boosted transactions, together with a significant spread between revenue captured in the primary auction and revenue generated through Kairos.

As we show below, the strongest empirical support is for the weakened-competition mechanism. After the transition to Kairos, the same searchers remain central, total value extraction changes only modestly, and primary auction revenue declines. By contrast, we find more limited evidence for unbundling or information-based efficiency gains.

\section{Methodology}
We now describe the datasets used in our analysis and the methodology for estimating arbitrage profits. Our primary data source is Dune Analytics~\cite{dune_analytics}.\footnote{The queries used in the analysis are publicly accessible: Timeboost auctions, \url{https://dune.com/queries/6219197}; CEX--DEX transactions, \url{https://dune.com/queries/6219209}; and Kairos payments, \url{https://dune.com/queries/6813491}.}

\subsection{Datasets}
Our datasets cover two complementary windows. The main auction and time-boosted transaction datasets span August 2025 through March 12, 2026. For the analysis of the Kairos transition, regular non-time-boosted CEX--DEX transactions and Kairos payments are collected over a focused February 1--March 12, 2026 window.

\noindent\textbf{Timeboost auctions.}
We collect Timeboost auction data from August 2025 until March 12, 2026. The dataset contains 298{,}530 auction rounds with at least one bid. We obtain detailed bidding data from public endpoints provided by Arbitrum~\cite{arbitrum_timeboost_historical_bids}. For each round, we observe the start and end time of the one-minute interval during which the auction winner holds fast-lane access. We use this interval to assign time-boosted transactions to auction rounds based on block time.

\noindent\textbf{Transactions.}
In total, we identify 30{,}310{,}245 time-boosted transactions executed during our study period. Of these, 27{,}399{,}487 (90.4\%) target contracts belonging to Wintermute\footnote{Wintermute CEX--DEX contracts: \texttt{0xcb43d843f6cadf4f4844f3f57032468aadd9b95c} and \texttt{0x27920e8039d2b6e93e36f5d5f53b998e2e631a70}.} and Selini\footnote{Selini CEX--DEX contract: \texttt{0xee2e7bbb67676292af2e31dffd1fea2276d6c7ba}.}. Wintermute accounts for 22{,}383{,}187 such transactions, while Selini accounts for 5{,}016{,}300. Since these contracts are associated with non-atomic CEX--DEX arbitrage, we apply a heuristic to identify transactions likely executing such strategies. We classify a transaction as CEX--DEX arbitrage if it targets these contracts, emits only one swap-related event, and trades a known liquid pair involving WETH, WBTC, USDC, or USDT. Using this heuristic, we label 11{,}320{,}741 Wintermute transactions and 3{,}399{,}555 Selini transactions as CEX--DEX arbitrage. These correspond to 50.6\% and 67.8\% of their respective time-boosted contract interactions. Applying the same heuristic to regular, non-time-boosted transactions targeting these contracts, we identify an additional 1{,}290{,}594 Wintermute and 316{,}927 Selini CEX--DEX arbitrage transactions during the February 1--March 12 analysis window.

\noindent\textbf{Kairos Payments.}
Using Arbitrum transaction traces, we identify all transactions during the February 1--March 12 analysis window in which Kairos's payment contract receives an explicit ETH transfer with positive value. In total, we identify 39{,}521 such payments.

\subsection{Markouts and Profits}
To estimate the PnL of Wintermute and Selini, we measure the realized value of their CEX--DEX arbitrage trades using markouts. The searchers' exact off-chain trading strategies are unobserved, including when and how positions accumulated on DEXes are offset on CEXes. Consider an arbitrage trade $i$ executed on-chain at time $t$ that purchases $x$ units of token A for $y$ units of token B on a DEX. Assuming the position is offloaded on a CEX after a markout horizon $m$ at prices $P_A(t+m)$ and $P_B(t+m)$, the profit is defined as
$$
\Pi(i)_{t,m}:=x\,P_A(t+m)-y\,P_B(t+m)-\text{fees}.
$$

Unless otherwise stated, our empirical analyses involving markouts restrict attention to trades with positive estimated PnL, thereby focusing on realized profitable arbitrage activity. For the markout-horizon sensitivity exercise, however, we report net PnL over a fixed sample of all identified trades, since applying the positive-PnL filter separately at each horizon would change the composition of the sample.

For our baseline specification, we use a markout horizon of $m=5$ seconds. This choice reflects a trade-off: very short horizons may value the DEX leg before prices have sufficiently adjusted across on-chain and off-chain markets, whereas longer horizons are increasingly likely to incorporate price movements unrelated to the original arbitrage opportunity.

To assess the sensitivity of the markout valuation to the chosen horizon, we recompute net markout PnL over a fixed sample of all identified time-boosted trades in August 2025 using horizons of $m\in\{1,2,5,10,30\}$ seconds. As shown in~\Cref{tab:markout_robustness_final} in the Appendix, estimated PnL increases between one and five seconds but is comparatively stable thereafter. Relative to the five-second baseline, total estimated PnL is only $1.2\%$ higher at ten seconds and $5.0\%$ higher at thirty seconds. At ten seconds, the corresponding differences are $1.8\%$ for Wintermute and $0.7\%$ for Selini. We therefore select five seconds as the shortest horizon in this relatively stable range, allowing time for cross-market price adjustment while limiting contamination from subsequent market movements.


We obtain price data from the Binance historical data repository~\cite{binance_market_data}, using USDT trading pairs sampled at one-second resolution granularity. Following~\cite{cex_dex}, we compute markouts using CEX mid-prices.

\subsection{Limitations}
Our methodology has several limitations.
\begin{enumerate}
    \item The heuristic used to detect CEX--DEX arbitrage captures only simple swaps involving a small set of liquid assets known to trade on CEXes. In practice, non-atomic arbitrage searchers may execute more complex strategies involving a long tail of assets~\cite{cex_dex}, as well as over-the-counter markets or venues on other chains~\cite{oz2025crosschainarbitragefrontiermev}.
    \item Profit estimation assumes that the entire inventory is unwound in a single trade at the chosen markout horizon. In reality, traders may unwind positions gradually to manage price risk, or may choose to retain inventory exposure.
    \item Trade coverage is limited by the DEX protocol data available on Dune Analytics.
    \item The observability of payments to Kairos is limited to on-chain transfers and does not include payments made through its CEX--DEX subscription model.
    \item We assume $1\text{USDT}=1\text{USDC}=1\text{USD}$, which may not always hold in practice.
\end{enumerate}
However, because the methodology is applied consistently across regimes, any resulting over- or underestimation of profits should affect all periods similarly and still allow meaningful comparisons across regimes.

\section{Empirical Results}
We now present the empirical results. We first evaluate auction effectiveness during the pre-transition competition phase, before turning to the emergence of Kairos-mediated JIT resale and its effects on bidding behavior, auction revenue, and surplus allocation.

\subsection{Auction Effectiveness in the Competition Phase}

We first assess the effectiveness of the AOT auction during the competition phase, before the Kairos transition. To do so, we measure the relationship between auction bids and estimated searcher PnL at the round level. \Cref{tab:bid_pnl_corr} shows that bids are noisy predictors of arbitrage profits. During the competition phase, the Pearson correlations between per-round profits and both the top bid and the paid bid are statistically significant but economically modest. Spearman correlations are higher for both searchers, suggesting that bidders could directionally identify higher-value intervals, but had limited ability to predict the magnitude of realized value.

\begin{table}[t!]
\centering
\caption{Pearson and Spearman correlations between per-round Timeboost auction bids and total positive-PnL time-boosted searcher profits during the competition phase. Auction bids are denominated in ETH, while searcher profits are estimated using a 5\,s markout. $n$ denotes the number of auction rounds.}
\label{tab:bid_pnl_corr}
\begin{tabular}{lcccc r}
\toprule
Winner & \multicolumn{2}{c}{Pearson} & \multicolumn{2}{c}{Spearman} & $n$ \\
\cmidrule(lr){2-3}\cmidrule(lr){4-5}
 & Top Bid & Paid Bid & Top Bid & Paid Bid & \\
\midrule
Wintermute & $0.321$ & $0.348$ & $0.449$ & $0.458$ & 108{,}489 \\
Selini     & $0.115$ & $0.109$ & $0.337$ & $0.305$ &  93{,}246 \\
\bottomrule
\end{tabular}
\end{table}

One way to interpret this pattern is through the relationship between arbitrage profits and return variance. Return variance is a moderately informative predictor of estimated arbitrage profits for the relevant trading pairs, but the fit is substantially stronger over longer horizons than over the one-minute intervals used in the Timeboost auction. Consistent with this, both the top bid and the paid bid are noisy predictors of short-run markout profits, but become more informative when profits are aggregated over longer forward windows. This suggests that bidders are not simply unable to identify valuable market conditions; rather, the main difficulty is pricing the realized value of a specific one-minute fast-lane interval ahead of time. This pattern is summarized in \Cref{tab:bid_pnl_horizons}.

\begin{table}[t!]
\centering
\small
\caption{Bid--PnL correlations at multiple horizons during the competition phase before the Kairos transition. Correlations are computed over forward windows of length $W$, where one auction round is approximately one minute. For each window, the bid variable is the mean top or paid bid over the window, and PnL is the total time-boosted PnL over the same window. The sample contains $n=201{,}737$ round-level observations.}
\label{tab:bid_pnl_horizons}
\begin{tabular}{lccccc}
\toprule
Bid type & 5 min & 10 min & 15 min & 30 min & 1 hour \\
\midrule
\multicolumn{6}{l}{\textit{Panel A: Spearman correlations}} \\
Top bid  & $+0.681$ & $+0.743$ & $+0.768$ & $+0.794$ & $+0.804$ \\
Paid bid & $+0.650$ & $+0.714$ & $+0.742$ & $+0.774$ & $+0.792$ \\
\addlinespace
\multicolumn{6}{l}{\textit{Panel B: Pearson correlations}} \\
Top bid  & $+0.195$ & $+0.245$ & $+0.278$ & $+0.342$ & $+0.401$ \\
Paid bid & $+0.178$ & $+0.215$ & $+0.236$ & $+0.316$ & $+0.378$ \\
\bottomrule
\end{tabular}
\end{table}

\begin{figure}[t!]
\centering
\includegraphics[width=\textwidth]{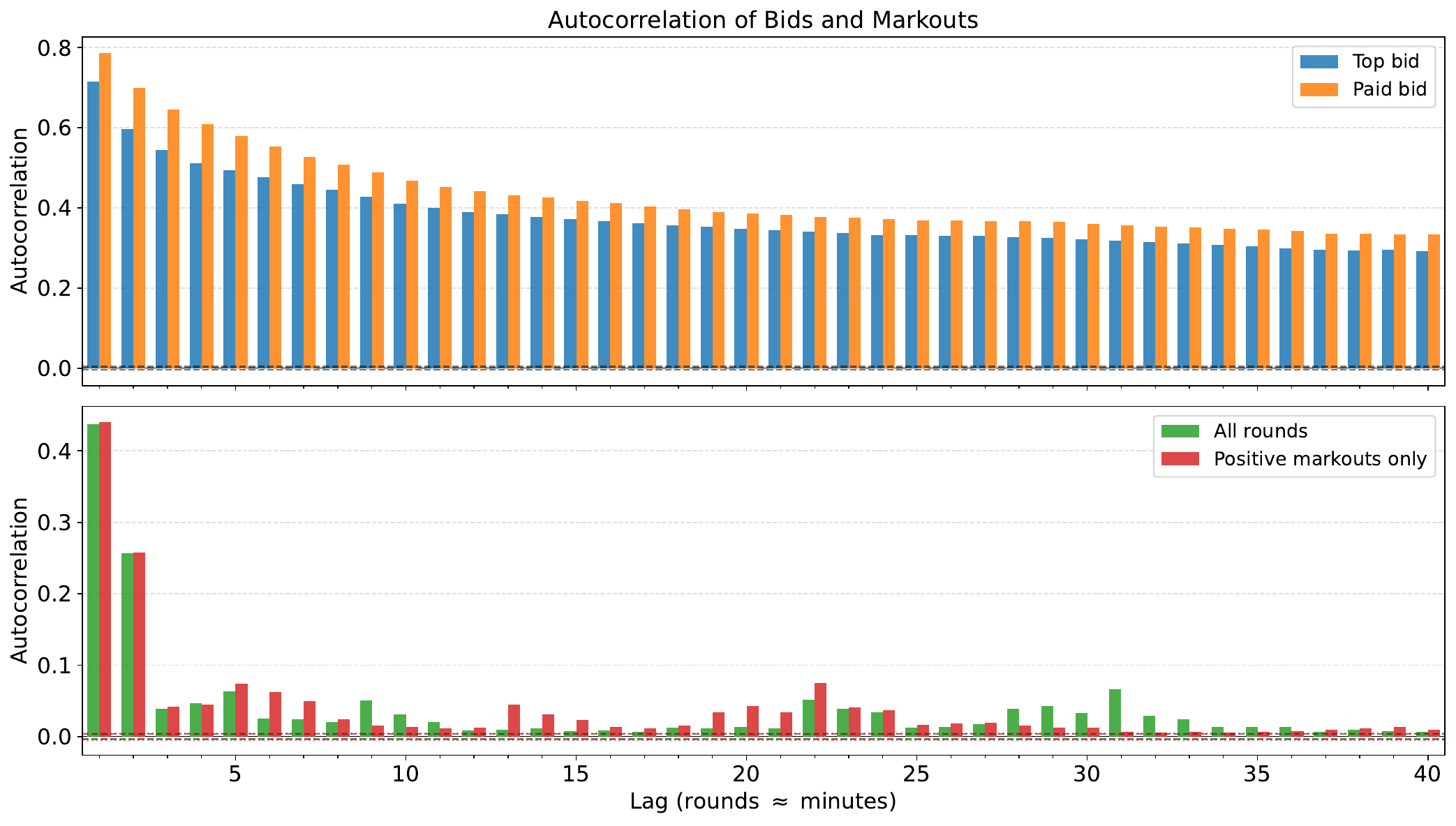}
\caption{Autocorrelation functions of auction bids and per-round markout PnL during the competition phase before the Kairos transition. \textit{Top:} Autocorrelation of the top and paid bids up to 40 lags, where one lag corresponds approximately to one auction round, or one minute. \textit{Bottom:} Autocorrelation of per-round time-boosted PnL for all rounds and for the positive-markout subsample. Dashed and dotted lines indicate 95\% confidence bands.}
\label{fig:aft_autocorr}
\end{figure}

Consistent with this interpretation, auction bids exhibit strong and slowly decaying autocorrelation during the competition phase, as shown in \Cref{fig:aft_autocorr}. This implies that bids reflect persistent market conditions rather than only round-specific realized opportunities. By contrast, per-round markout PnL is much less persistent. The all-round markout series exhibits little autocorrelation beyond the first two lags, while positive-markout rounds show only limited short-run clustering. This difference is consistent with the idea that bidders can track broader market conditions, but face substantial uncertainty about the realized value of any individual one-minute interval.

\begin{figure}[t!]
\centering
\includegraphics[width=\textwidth]{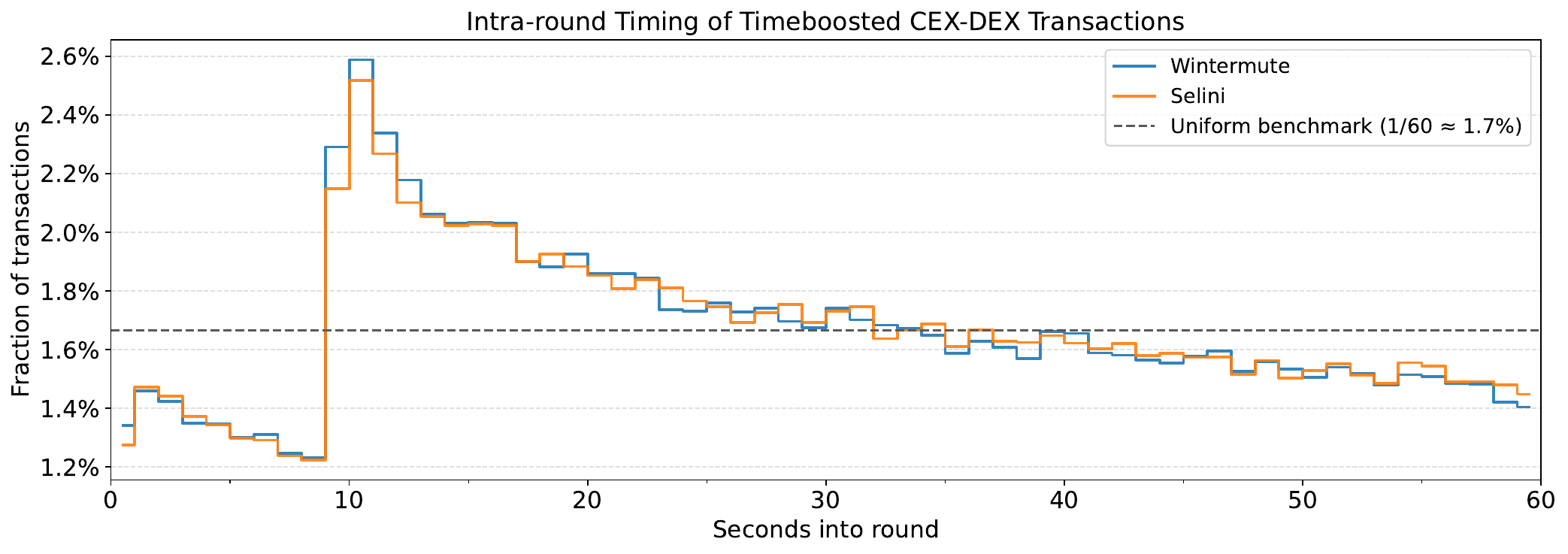}
\caption{Intra-round timing distribution of time-boosted CEX--DEX transactions by Wintermute and Selini during the competition phase. Each line represents the fraction of transactions submitted in a given second of the one-minute Timeboost round. Dashed line indicates the uniform distribution benchmark.}
\label{fig:intraslot_timing}
\end{figure}

The intra-round timing distribution motivates a shorter-window analysis. As shown in \Cref{fig:intraslot_timing}, time-boosted CEX--DEX transactions are not uniformly distributed within the round. Transaction activity initially declines and then exhibits a pronounced spike around second 10 of the Timeboost interval, which corresponds approximately to the beginning of the next wall-clock minute. Because we do not observe the searchers' trading logic or off-chain opportunity signals, we cannot identify the cause of this spike directly.

One possible interpretation is clock-time scheduling. Hasbrouck and Saar~\cite{hasbrouck_low-latency_2013} document periodicities in Nasdaq message activity associated with automated strategies that access the market at fixed wall-clock intervals. The alignment of the spike with the minute boundary is consistent with a similar mechanism. However, it could
also reflect periodic updates on external venues, protocol-specific timing effects, or common reactions by the dominant searchers to the same market signals. We therefore interpret the timing pattern as suggestive rather than causal.

Because activity is concentrated later in the interval, the first seconds of each round contribute disproportionately little to total realized PnL, which is relevant when interpreting correlations measured over shorter markout windows.
We therefore repeat the round-level correlation exercise using shorter markout windows. Instead of measuring PnL over the full one-minute Timeboost interval, we compute markouts over the first 15 and 30 seconds of the interval. This test captures competing forces. On the one hand, bidders may be better informed about more immediate opportunities, which could increase the correlation between bids and realized PnL. On the other hand, shorter windows contain less aggregation, so realized profits are likely to be noisier than over the full minute. Moreover, because activity is concentrated later in the round, shorter windows also mechanically capture less of the realized opportunity, which works against finding strong correlations. The results in \Cref{tab:short_markout_corr} support the latter interpretations: correlations remain positive but economically modest, and are generally weaker than for longer aggregation windows.


\begin{table}[t!]
\centering
\small
\caption{Correlations between auction bids and positive-PnL time-boosted markouts over shorter horizons during the competition phase before the Kairos transition. Markouts are computed over the first 15 and 30 seconds of the one-minute Timeboost interval. $n$ denotes the number of auction rounds.}
\label{tab:short_markout_corr}
\begin{tabular}{lccccc}
\toprule
Winner & \multicolumn{2}{c}{15 seconds} & \multicolumn{2}{c}{30 seconds} & $n$ \\
\cmidrule(lr){2-3}\cmidrule(lr){4-5}
 & Top bid & Paid bid & Top bid & Paid bid & \\
\midrule
\multicolumn{6}{l}{\textit{Panel A: Pearson correlations}} \\
Wintermute & $+0.240$ & $+0.276$ & $+0.282$ & $+0.314$ & 116{,}679 \\
Selini     & $+0.087$ & $+0.098$ & $+0.101$ & $+0.110$ & 133{,}774 \\
\addlinespace
\multicolumn{6}{l}{\textit{Panel B: Spearman correlations}} \\
Wintermute & $+0.385$ & $+0.386$ & $+0.421$ & $+0.424$ & 116{,}679 \\
Selini     & $+0.292$ & $+0.277$ & $+0.315$ & $+0.302$ & 133{,}774 \\
\bottomrule
\end{tabular}
\end{table}

\subsection{Bid Coordination and Just-in-Time Resale}
In this section, we analyze the transition from direct competition in the primary Timeboost auction to Kairos-mediated fast-lane access. We focus on how this shift reshapes bidding behavior, primary auction revenue, and the distribution of surplus.

\subsubsection{Timeline}\label{sec:timeline}
Within our study period, we identify three key on-chain events that delineate the observed market dynamics:

\begin{enumerate}
    \item \textbf{2026-02-12 20:31:51 UTC} --- Wintermute and Selini cease competitive bidding in the primary Timeboost auction and begin sourcing fast-lane access through Kairos.

    \item \textbf{2026-02-18 20:01:51 UTC} --- Arbitrum raises the Timeboost reserve price from 0.001 ETH to 0.0075 ETH.

    \item \textbf{2026-02-25 19:49:51 UTC} --- Arbitrum lowers the reserve price back to 0.001 ETH.
\end{enumerate}

These events divide the February 1--March 12, 2026 transition-analysis window into four sub-periods: \textit{Competition} (pre-transition February), \textit{Kairos} (February 12--18), \textit{Reserve Price Adaptation} (February 18--25), and \textit{Steady State} (after February 25). The label \textit{Competition} refers specifically to the February segment of the broader pre-transition competition phase. We use this shorter pre-transition window because auction dynamics in February are representative of the full pre-transition period: the August 2025--February 2026 data exhibit a stable structure (see \Cref{fig:auction_winners}).

\subsubsection{Bidding Behavior}\label{sec:bidding_behavior}
During the coordination phase, we observe 47{,}689 Timeboost auctions, of which Kairos (54.5\%), Selini (23.6\%), and  Wintermute (21.6\%) together account for 99.7\% of all wins (see~\Cref{tab:rounds_won}). In total, 117{,}146 bids were placed, with an average of 2.46 bids per round across 14 distinct bidding addresses. Selini accounts for the highest share of bids (38.7\%), followed by Wintermute (32.7\%) and Kairos (25.0\%); together these three entities are responsible for 96.4\% of all bids. We summarize
auction participation and active periods for each bidding address in
\Cref{tab:unique_bidders} in the Appendix. We interpret these as distinct on-chain bidding accounts and do not assume that unattributed addresses necessarily correspond to distinct economic entities.

\begin{table}[t!]
\centering
\caption{Share of auction rounds won by each entity per period.}
\label{tab:rounds_won}
\resizebox{\textwidth}{!}{%
\begin{tabular}{lrrrrr}
\toprule
Entity
  & Competition
  & Kairos
  & Res.\ Price Adapt.
  & Steady State
  & Overall \\
\midrule
Wintermute & 45.0\% (7{,}682/17{,}065) &  0.0\% (1/8{,}593)      & 72.7\% (141/194) & 11.3\% (2{,}478/21{,}837) & 21.6\% (10{,}302/47{,}689) \\
Selini     & 53.4\% (9{,}109/17{,}065) &  0.2\% (18/8{,}593)     &  3.6\% (7/194)   &  9.8\% (2{,}134/21{,}837) & 23.6\% (11{,}268/47{,}689) \\
Kairos     &  1.4\% (244/17{,}065)     & 99.6\% (8{,}559/8{,}593) & 23.7\% (46/194)  & 78.4\% (17{,}119/21{,}837) & 54.5\% (25{,}968/47{,}689) \\
\bottomrule
\end{tabular}}
\end{table}

Looking only at the number of bids per round might suggest that the competitiveness of the auction was not significantly affected by the adoption of Kairos. \Cref{tab:bidder_combinations} shows that the number of rounds in which all three entities placed a bid is highest after this transition. However, examining the gap between the winning bid (top bid) and the paid bid reveals a clear decline in auction competitiveness.


\begin{table}[t!]
\centering
\caption{Share of auction rounds by bidder participation combination in each period.
WM = Wintermute, Sel = Selini, Kai = Kairos.}
\label{tab:bidder_combinations}
\resizebox{\textwidth}{!}{%
\begin{tabular}{lrrrrr}
\toprule
Combination
  & Competition
  & Kairos
  & Res.\ Price Adapt.
  & Steady State
  & Overall \\
\midrule
WM + Sel + Kai  &  1.5\% (262/17{,}065)    & 10.9\% (935/8{,}593)    &  1.5\% (3/194)     & 85.3\% (18{,}621/21{,}837) & 41.6\% (19{,}821/47{,}689) \\
WM + Sel        & 88.8\% (15{,}161/17{,}065) &  0.0\% (0/8{,}593)    &  4.1\% (8/194)     &  5.1\% (1{,}118/21{,}837)  & 34.2\% (16{,}287/47{,}689) \\
WM + Kai        &  0.0\% (4/17{,}065)      &  0.2\% (14/8{,}593)    & 10.3\% (20/194)    &  6.7\% (1{,}471/21{,}837)  &  3.2\% (1{,}509/47{,}689)  \\
Sel + Kai       &  0.2\% (41/17{,}065)     & 87.1\% (7{,}485/8{,}593) &  0.5\% (1/194)   &  1.2\% (257/21{,}837)      & 16.3\% (7{,}784/47{,}689)  \\
Only WM         &  1.2\% (198/17{,}065)    &  0.0\% (0/8{,}593)     & 67.5\% (131/194)   &  1.4\% (314/21{,}837)      &  1.3\% (643/47{,}689)      \\
Only Sel        &  8.2\% (1{,}399/17{,}065) &  0.0\% (4/8{,}593)    &  3.1\% (6/194)     &  0.2\% (35/21{,}837)       &  3.0\% (1{,}444/47{,}689)  \\
Only Kai        &  0.0\% (0/17{,}065)      &  1.8\% (155/8{,}593)   & 12.9\% (25/194)    &  0.1\% (20/21{,}837)       &  0.4\% (200/47{,}689)      \\
None            &  0.0\% (0/17{,}065)      &  0.0\% (0/8{,}593)     &  0.0\% (0/194)     &  0.0\% (1/21{,}837)        &  0.0\% (1/47{,}689)        \\
\bottomrule
\end{tabular}}
\smallskip
\begin{minipage}{\textwidth}
\end{minipage}
\end{table}

\begin{figure}[t!]
    \centering
    \includegraphics[width=\linewidth]{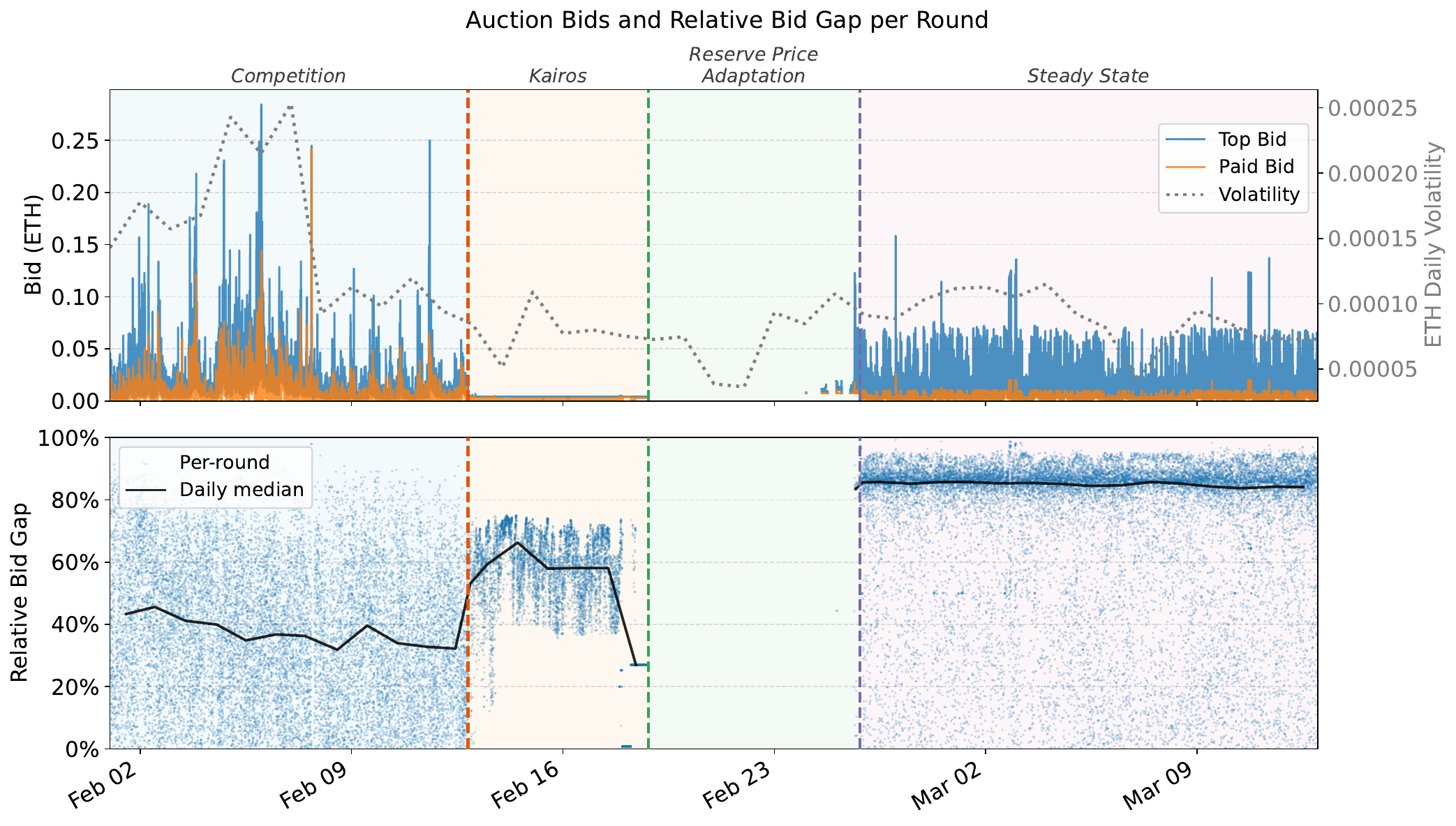}
    \caption{Timeboost auction bidding dynamics during the transition-analysis window. \textit{Top:} Top bid and paid bid per auction round, measured in ETH. The dotted line shows daily ETH volatility on the right axis. \textit{Bottom:} Relative bid gap between the top bid and the paid bid for each auction round, shown as points, together with the daily median, shown as a line.  Breaks in the daily-median line indicate periods with no auction rounds receiving bids.}
    \label{fig:bids_and_gap}
\end{figure}

\Cref{fig:bids_and_gap} shows that during the \textit{Competition} phase, both the top bid and the paid bid move closely with volatility in ETH prices\footnote{We compute volatility as the standard deviation of log returns from consecutive one-second ETH/USDT mid-price observations.}. This relationship is confirmed by the correlation measures reported in \Cref{tab:vol_corr_eth} in the Appendix, which show strong and statistically significant linear relationships between volatility and both paid and top bids. The absolute bid gap also increases with volatility, indicating that while both bids respond to more valuable trading opportunities, the top bid tends to rise more strongly. During this period, the median daily relative bid gap is 37.3\%.

Once Wintermute and Selini retreat from the auction and begin sourcing fast-lane access through Kairos (the \textit{Kairos} period), the median daily relative bid gap increases to 57.5\%. At the same time, the paid bid only weakly tracks price volatility, while the top bid shows no statistically significant relationship. During this period, Kairos wins 99.6\% of the rounds and predominantly submits a fixed bid of 0.004 ETH, while the paid price typically remains only marginally above the reserve price, usually submitted by Selini (see~\Cref{tab:bidder_combinations}). As a result, the bid gap becomes negatively correlated with volatility, which can be explained by the paid price adjusting slightly with volatility while the top bid remains largely fixed. Consequently, auction pricing in this regime remains largely compressed despite changing market conditions.

One notable deviation during the \emph{Kairos} period occurs in the final 24 hours before the reserve-price increase. At 20:00:51 UTC on February 17, the previously inactive and unattributed address \texttt{0x3141592601bf441700f150bef82db6040f4fe373} begins submitting several different bid amounts. After approximately two hours, it switches to a fixed bid of 0.003970 ETH, only 0.75\% below Kairos's predominant fixed bid of 0.004 ETH, and submits this amount in every round from 23:00 on February 17 until approximately 06:00 on February 18. Because Timeboost is a second-price auction, the address becomes the price-setting second-highest bidder in most of these rounds, sharply reducing the relative bid gap and increasing the price paid by Kairos.

From 06:00 until the reserve-price update approximately 14 hours later, the address instead submits a fixed bid of 0.002921 ETH, producing the roughly 27\% relative-bid-gap plateau visible in \Cref{fig:bids_and_gap}. The address wins only 13 rounds, approximately 1\% of the rounds in which it participates, but sets the paid bid in
1{,}238 rounds, or 96\% of its active rounds. In the remaining rounds, another address submits the second-highest bid. The address is listed in \Cref{tab:unique_bidders}, but we find no public attribution for it.

During the \emph{Reserve Price Adaptation} period, the reserve price increases from 0.001 ETH to 0.0075 ETH and bidding activity nearly disappears. Only 194 rounds receive at least one bid, the majority of which (131 rounds) involve Wintermute as the sole bidder. Among these active rounds, bidding activity is again positively correlated with price volatility, consistent with Wintermute deriving value from external price movements as a CEX--DEX searcher. The apparent increase in the daily median relative bid gap in \Cref{fig:bids_and_gap} should therefore not be interpreted as a gradual change in bidding behavior, as it is based on a small number of active rounds. In rounds with only one bidder, the paid price is determined by the reserve, so the relative bid gap reflects the difference between Wintermute's bid and the elevated reserve price. Bidding resumes more consistently only after the reserve price returns to 0.001 ETH.

Finally, the market enters what we refer to as the \textit{Steady State} once Arbitrum lowers the reserve price back to its original value of 0.001 ETH. Despite having the same reserve price as the earlier \textit{Kairos} period, bidding behavior differs substantially. All three entities submit bids in 85.3\% of rounds, while Kairos wins 78.4\% of rounds. Both the top bid and the paid bid again exhibit strong positive correlations with volatility, as does the absolute bid gap. However, the paid bid remains effectively capped in most rounds, as reflected in \Cref{fig:bids_and_gap}. As a result, the top bid rises more strongly than the paid bid when trading opportunities become more valuable, and the median relative bid gap stabilizes at 85.2\%. Thus, the paid bid amounts to only approximately 14.8\% of the top bid, indicating that the primary auction captures only a limited share of the value reflected in submitted bids.

Unlike in the preceding \textit{Kairos} period, Kairos no longer submits a predominantly fixed bid or wins nearly every auction. Across the 21{,}837 rounds with at least one bid, Kairos participates in 93.3\% and wins 84.0\%. The median relative gap between Kairos's winning bid and the second-highest bid is approximately 86.0\%, whereas it falls to 32.0\% in rounds that Kairos loses. The difference is even more pronounced in absolute terms: the median gap is approximately 0.015 ETH when Kairos wins, compared with 0.001 ETH when another bidder wins.

\begin{figure}[t!]
    \centering
    \includegraphics[width=\linewidth]{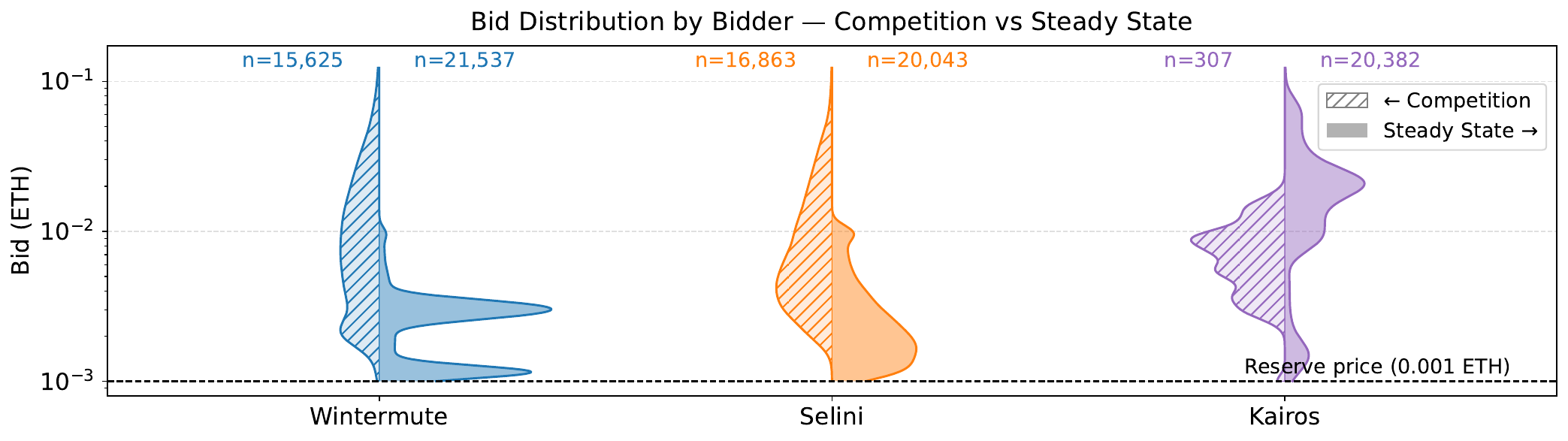}
    \caption{Bid distributions for Wintermute, Selini, and Kairos. The left half of each violin corresponds to the \textit{Competition} period, while the right half corresponds to the \textit{Steady State}. The dashed horizontal line indicates the reserve price (0.001 ETH).}
    \label{fig:bid_distribution_violin}
\end{figure}

This asymmetry is also reflected in the bid distributions shown in \Cref{fig:bid_distribution_violin}. During the \textit{Competition} period, median bids are relatively close across Wintermute, Selini, and Kairos (0.0075, 0.0062, and 0.0069 ETH, respectively). In the \textit{Steady State}, they diverge substantially: Kairos's median bid rises to 0.0182 ETH, whereas the corresponding medians for Wintermute and Selini fall to 0.0027 and 0.0021 ETH. The same pattern appears in the upper tail of the distributions (p99: Wintermute 0.0096 ETH, Selini 0.0101 ETH, and Kairos 0.0703 ETH). A more detailed distribution is reported in \Cref{tab:bid_distribution} in the Appendix. These patterns show that the renewed participation of Wintermute and Selini does not constitute a return to the competitive bidding observed before the adoption of Kairos. Kairos bids substantially more aggressively, while the other two searchers generally submit bids close to the reserve.

\begin{figure}[t!]
    \centering
    \includegraphics[width=\linewidth]{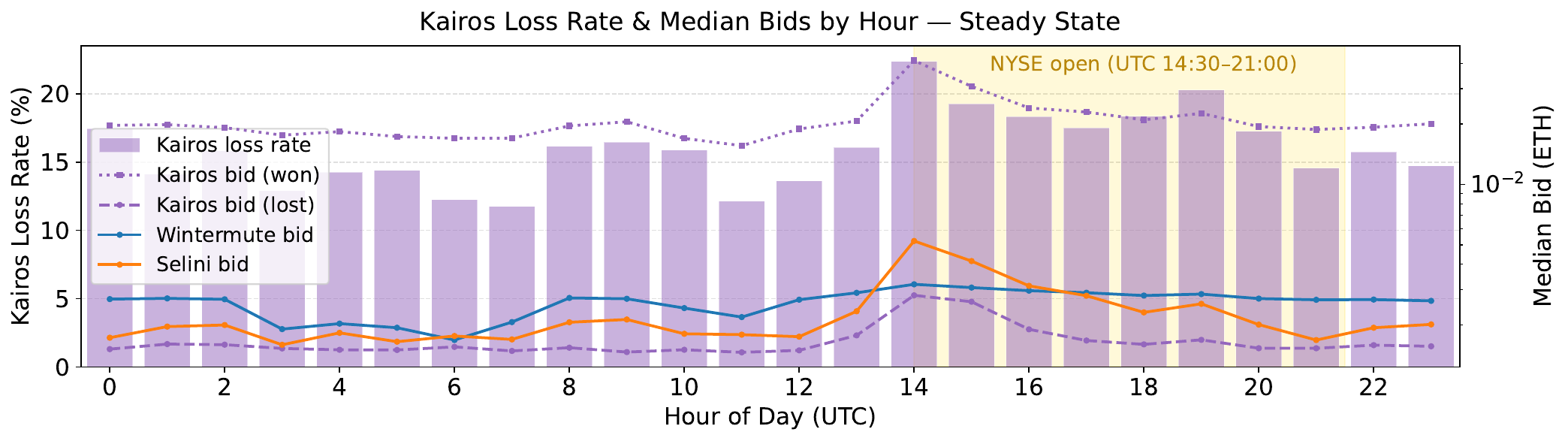}
    \caption{Kairos loss rate and median bids by hour of day in the \textit{Steady State} period. The shaded region indicates UTC hour bins overlapping the regular NYSE session (14{:}30--21{:}00 UTC during EST).}
    \label{fig:kairos_loss}
\end{figure}

To better understand when Kairos loses in the \textit{Steady State}, we examine the temporal distribution of its losses. As shown in \Cref{fig:kairos_loss}, Kairos loses more frequently during U.S.\ market hours, particularly around the NYSE opening: its loss rate is 19.0\% during NYSE hours, compared with 14.5\% outside those hours. Selini, in particular, submits somewhat higher bids during these periods. Nevertheless, Kairos's median bid in rounds it loses remains substantially below its median winning bid. Heightened market activity therefore explains only part of the variation in Kairos's losses. More generally, losses tend to occur when Kairos submits a relatively low bid or does not participate, rather than when Wintermute or Selini submit bids comparable to Kairos's usual winning bids.

One possible interpretation is that the temporary reserve-price increase altered participants' expectations about future protocol responses, making a complete withdrawal from the primary auction less attractive after the reserve returned to 0.001 ETH. This could help explain why all three entities maintain visible bidding activity in the \textit{Steady State}, while Kairos submits higher and more variable bids to preserve its access to the fast-lane. Because participants' expectations and off-chain arrangements are unobserved,
we cannot identify the cause of this shift directly. We therefore interpret the \textit{Steady State} as a modified coordination regime with limited residual competition, rather than a return to the earlier competitive environment.

\subsubsection{Searcher Activity}\label{sec:searcher_activity}
\begin{figure}[t!]
    \centering
    \includegraphics[width=\linewidth]{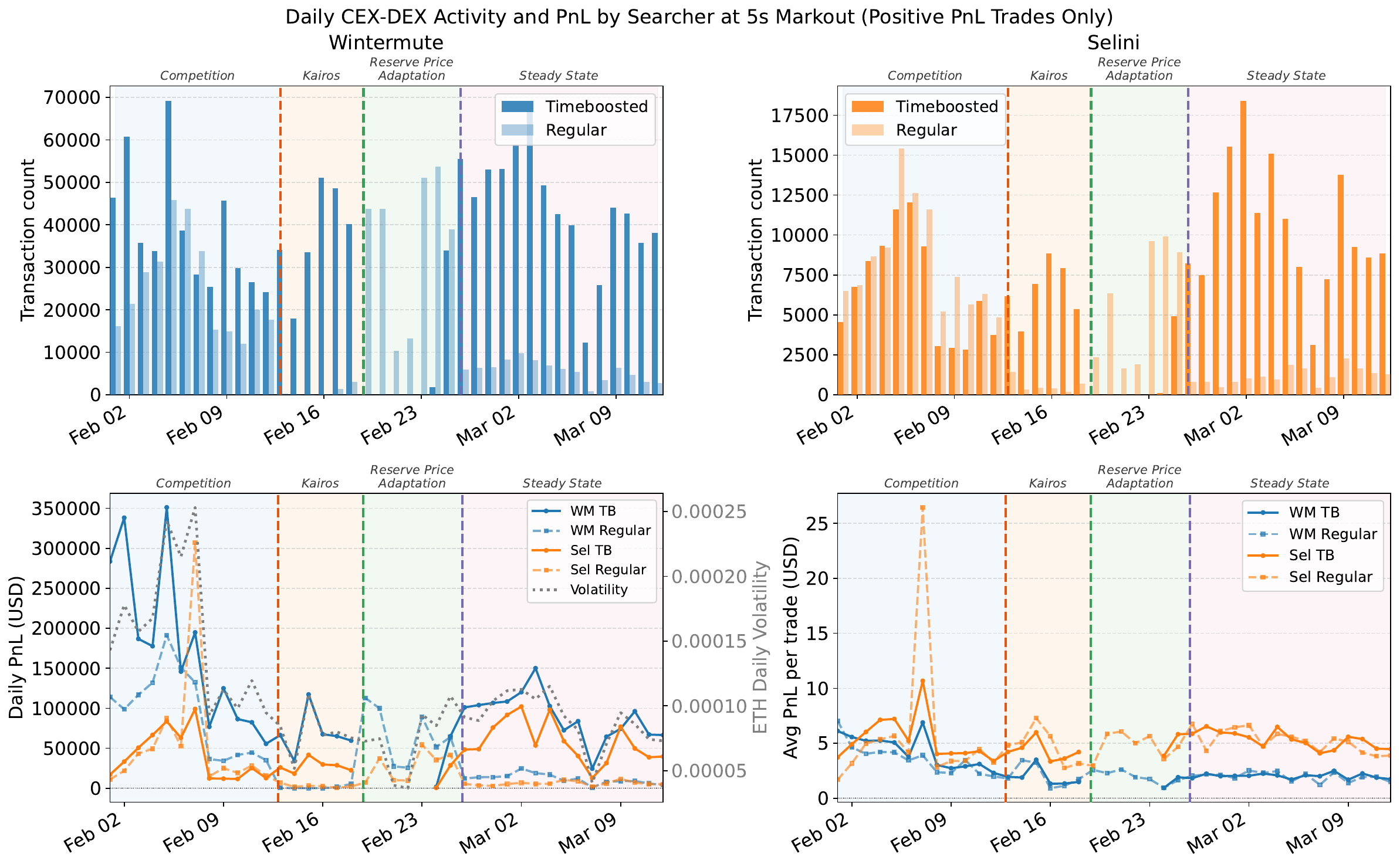}
    \caption{CEX--DEX arbitrage activity by Wintermute and Selini during the transition-analysis window. 
    \textit{Top-left:} time-boosted and regular transaction count by Wintermute. 
    \textit{Top-right:} time-boosted and regular transaction count by Selini. 
    \textit{Bottom-left:} daily PnL from time-boosted and regular transactions for Wintermute and Selini. The dotted line shows daily ETH volatility (right axis). 
    \textit{Bottom-right:} average PnL per trade for time-boosted and regular transactions for Wintermute and Selini.}
    \label{fig:daily_activity}
\end{figure}

We next examine how the changing auction dynamics affect searcher activity and the use of the Timeboost fast-lane versus the regular-lane. The top-left and top-right panels in \Cref{fig:daily_activity} show the number of time-boosted and regular CEX--DEX transactions submitted over time by Wintermute and Selini, respectively. The bottom-left panel reports daily PnL, while the bottom-right panel shows the average PnL per trade per day for time-boosted and regular transactions. Consistent with our baseline PnL specification, we restrict attention to trades with positive estimated PnL.~\Cref{tab:pnl_summary_zone} in the Appendix summarizes the results.

Before the adoption of Kairos, in the \textit{Competition} phase, searchers submit CEX--DEX transactions through both the Timeboost fast-lane and the regular-lane when they do not control the fast-lane. Their usage patterns differ meaningfully: Wintermute earns more total PnL from time-boosted trades and submits relatively more of them, whereas Selini earns more total PnL from regular-lane trades and also achieves higher average profit per trade there. More broadly, although Wintermute trades more frequently and earns higher total profits overall, Selini tends to realize higher average profit per trade. These differences remain largely stable across auction regimes and suggest that the two searchers follow somewhat different trading models. At the same time, there are periods in which average profit per trade is similar across lanes, and sometimes even higher in the regular-lane for both searchers, indicating that Timeboost is not reserved only for the most profitable opportunities but also helps secure smaller trades that might otherwise be lost to latency competition. For both searchers, PnL from both time-boosted and regular-lane trades remains positively correlated with volatility (see~\Cref{tab:pnl_vol_corr} in the Appendix), consistent with the idea that higher volatility creates more opportunities to exploit stale DEX prices relative to rapidly moving CEX prices.

Once Wintermute and Selini begin using Kairos, both almost completely stop submitting transactions through the regular-lane, as they can now access the fast-lane via Kairos. This indicates that Kairos effectively substitutes for direct participation in the Timeboost auction, allowing searchers to retain fast-lane access without bidding competitively in the primary auction. During this \textit{Kairos} phase, the PnL from time-boosted trades remains strongly correlated with volatility for both searchers, whereas the relationship is considerably weaker for regular-lane trades, suggesting a clear shift in how the two submission lanes are used.

During the \textit{Reserve Price Adaptation} phase, when bidding activity was minimal and resumed only toward the end, both searchers mainly continued operating through the regular transaction lane. Their average PnL remains similar to the \textit{Kairos} period, but the volatility relationships appear to reverse: regular-lane transactions now show a strong correlation with volatility for both searchers, while the much less frequent time-boosted trades exhibit a weaker relationship.

Finally, once the reserve price returns to its original level, both searchers move back to the fast-lane while submitting relatively few regular-lane transactions. In this \textit{Steady State} period, both execute significantly more time-boosted trades relative to regular-lane trades than in the \textit{Competition} period, reflecting improved access to the fast-lane through Kairos. 

Overall, the adoption of Kairos allows searchers to access the fast-lane almost continuously, largely eliminating the need to rely on the slower regular-lane while time-boosted activity continues to track price movements closely. However, consistent with the weaker bidding competition documented above, this increased reliance on the fast-lane does not translate into a proportional increase in primary-auction revenue.

\subsubsection{Surplus}
Lastly, we examine how the changing auction dynamics affect the total surplus generated and its distribution across the involved parties. We define net surplus as
$$
\text{Net Surplus} := \sum_{\text{Wintermute, Selini}}\mathrm{PnL} - \text{transaction fees} - \text{auction bids paid},
$$
where transaction fees and auction bids paid together represent revenue captured by Arbitrum, while total searcher PnL captures the value generated by CEX--DEX arbitrage opportunities and ultimately shared among participants in the Timeboost ecosystem.

\begin{figure}[t!]
    \centering
    \includegraphics[width=\linewidth]{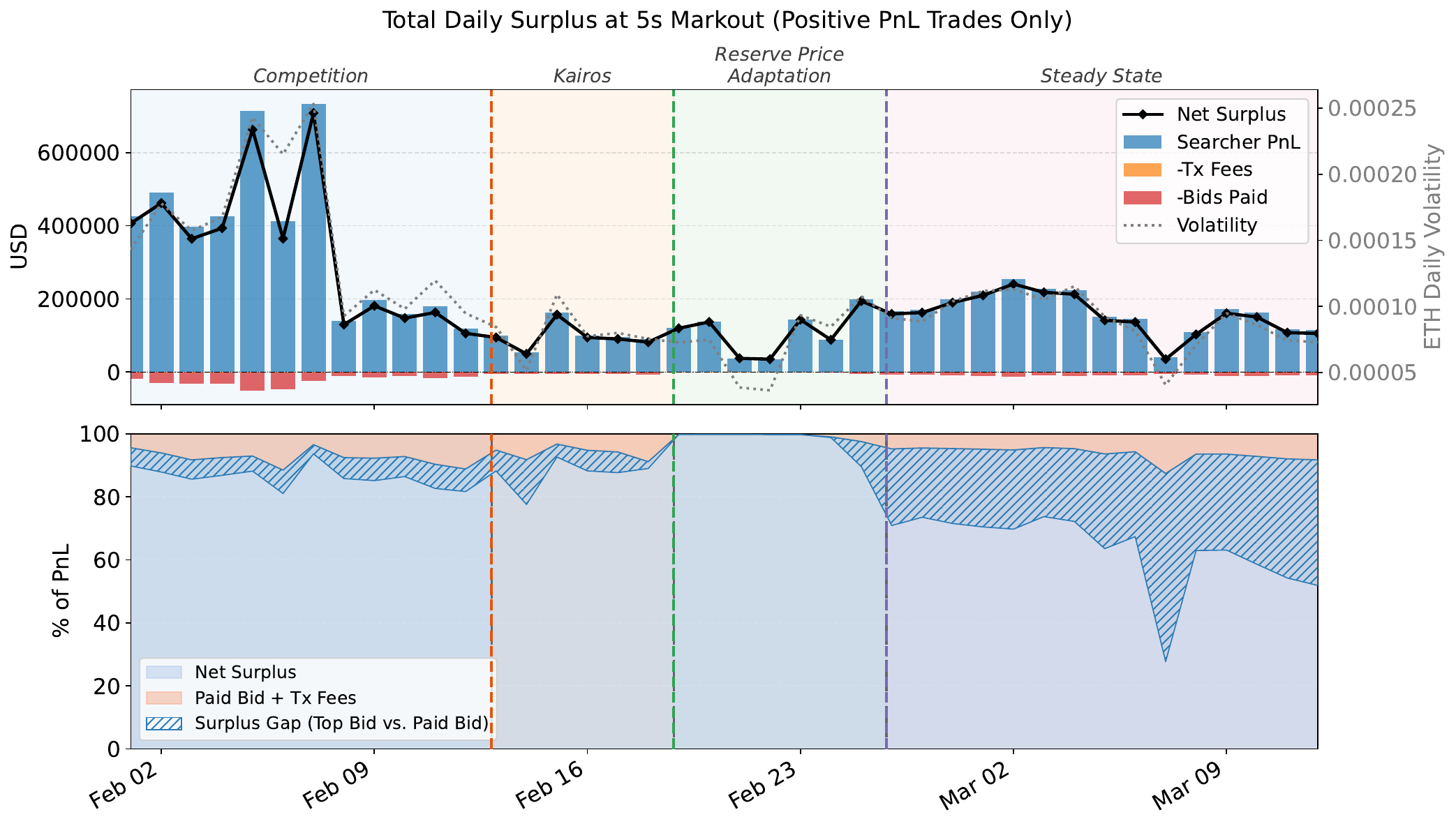}
    \caption{Surplus dynamics during the February 1--March 12, 2026 transition-analysis window.
    \textit{Top:} daily surplus decomposition into searcher PnL, transaction fees, and Timeboost auction bids paid. The black solid line shows the net surplus, while dotted line shows daily ETH volatility (right axis).
    \textit{Bottom:} share of searcher PnL retained as net surplus versus captured by Arbitrum through transaction fees and paid auction bids. The hatched area shows the difference between the share of surplus reflected in the top bid and the share captured through the paid bid.}
    \label{fig:total_surplus}
\end{figure}

\Cref{fig:total_surplus} presents the evolution of total surplus over time (top panel) and the share of searcher PnL captured by Arbitrum through auction bids and transaction fees (bottom panel). Throughout the study period, searcher PnL is strongly positively correlated with daily ETH volatility ($r = +0.96$, $p < 0.001$), with visible spikes during the \textit{Competition} phase when volatility was highest. Importantly, total searcher PnL remains of similar magnitude across regimes, suggesting that the observed changes in primary-auction competition mainly reflect a redistribution of surplus rather than an increase in total value generated.

We find that bids paid correspond on average to 7.4\% of total searcher PnL during the \textit{Competition} phase\footnote{This share is likely understated, as we restrict attention to positive-PnL trades. Including loss-making trades would reduce net searcher PnL and thus mechanically increase the captured share.}, but this declines following the adoption of the Kairos secondary market (\textit{Kairos}: 5.8\%, \textit{Reserve Price Adaptation}: 0.4\%, \textit{Steady State}: 6.0\%). This decline is even more pronounced when focusing only on time-boosted surplus and excluding regular-lane transactions (\textit{Competition}: 12.5\%, \textit{Steady State}: 6.7\%). This pattern is consistent with increasing reliance on the fast-lane via Kairos together with weaker direct competition in the primary auction.

Although bids in the Timeboost auction need not directly reveal true valuations because of the common-value component and strategic shading, stronger competition in the \textit{Competition} period likely allowed a larger share of surplus to be captured through the auction than in the \textit{Steady State}, where Kairos is often the only competitive participant with a high value for winning the lane. Consistent with this interpretation, the top bid corresponds to a substantially larger share of surplus in the final regime (\textit{Competition}: 13.6\%, \textit{Steady State}: 36.3\%), while the gap between the surplus reflected in the top bid and the surplus captured through the paid bid also widens markedly (bottom panel of~\Cref{fig:total_surplus}). The key change, therefore, is not that bidding ceases to respond to valuable trading opportunities, but that weaker competition causes a smaller fraction of the underlying value to be captured through the paid bid. Whether this reflects more aggressive shading, coordination, or lower standalone value from controlling the lane for bidders other than Kairos, the implication is the same: weaker competition increases the disconnect between value and revenue.

\begin{figure}[t!]
    \centering
    \includegraphics[width=\linewidth]{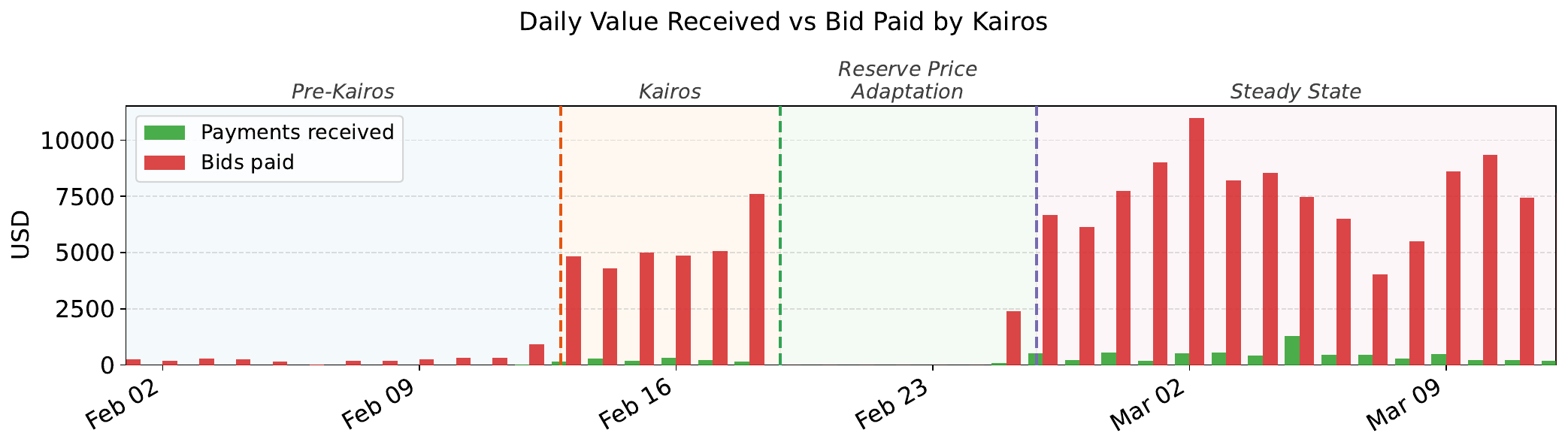}
    \caption{Kairos primary auction payments and resale revenue during the February 1--March 12, 2026 transition-analysis window. The figure compares the daily Timeboost auction bids paid by Kairos to the on-chain payments Kairos receives from transactions using its service.}
    \label{fig:kairos_payments}
\end{figure}

We now examine payments to Kairos to better understand how surplus may be distributed outside the primary auction. \Cref{fig:kairos_payments} compares the total daily bids paid by Kairos in the Timeboost auction with the on-chain payments it receives from transactions using its service. While Kairos's auction payments increase substantially once it becomes the dominant auction participant, the observed on-chain payments it receives do not increase proportionally.

Over the study period, Kairos pays a total of 151{,}302 USD in auction bids, while the on-chain payments we observe sum to only 8{,}001 USD. The resulting gap of more than 143{,}000 USD suggests that a substantial portion of the payments to Kairos likely occurs through alternative mechanisms that are not visible on-chain, such as its subscription model for CEX--DEX arbitrage searchers.

During periods when Kairos controls the fast-lane, Wintermute and Selini submit a combined total of 1{,}004{,}876 time-boosted CEX--DEX transactions with positive PnL, generating approximately 2{,}541{,}351 USD. However, none of the identified on-chain payments to Kairos originate from their transactions, suggesting that these entities primarily rely on Kairos's off-chain payment structure.

Taken together, these findings are consistent with the evidence above that, once Kairos becomes dominant, a smaller share of the value generated around Timeboost is captured through the primary auction. However, because payments to Kairos are only partially observable on-chain, the ultimate division of the remaining surplus between the intermediary and the searchers remains unclear. Hence, we cannot establish the extent using Kairos benefits Wintermute and Selini as the final surplus allocation is unobservable on-chain.

\section{Concluding Discussion}
We conclude by discussing what our empirical results imply for value extraction in AOT auctions. We first consider the effectiveness of AOT auctions when access is allocated through the primary auction alone, then examine how JIT resale changes revenue capture and surplus allocation. Finally, we discuss broader implications for Ethereum.

\subsection{Value Extraction Through Ahead-of-Time Auctions}
Analyzing Timeboost auction data in the competition phase shows that estimating immediate future arbitrage value is difficult. Bids are reasonably informative about longer-horizon average arbitrage profits, but they are noisy predictors of one-minute or shorter-horizon profits. This finding casts doubt on whether AOT auctions are effective MEV extraction mechanisms for rollups or other blockchains, even when no JIT resale market forms.

In a pure common-value auction with risk-neutral bidders, noisy ex-ante signals do not necessarily imply low expected seller revenue: bidders should condition on the winner's curse and shade their bids relative to their private signals, but competition can still transfer a substantial share of expected value to the seller. In practice, however, the risk-neutrality assumption might not hold perfectly. Bidders may be risk averse, face capital or inventory constraints, or differ in their private ability to exploit opportunities. Such private-value components weaken the link between common arbitrage value and bids. They also imply that selling ordering rights AOT may fail to allocate the fast-lane to the bidder with the highest realized value for it.

\subsection{Value Extraction if There is a Just-in-Time Resale Market}
The adoption of the Kairos resale market effectively dampens competition in Arbitrum's Timeboost auction, consistent with the weakened-competition channel being the dominant mechanism---though we find some limited supporting evidence for un-bundling and information effects as well. As a result, the protocol appears to capture a substantially smaller share of the value generated around Timeboost, while a larger share remains outside the primary auction. However, the exact distribution of this additional surplus between the searchers and Kairos remains unclear, as the payments associated with the resale mechanism are only partially observable on-chain.

The success of the Kairos resale market is also reflected in the increasing volume of time-boosted transactions once Kairos wins most auction rounds. However, this effect is not clearly visible in PnL, which is a noisier signal. As a result, the unbundling effect appears to be second-order relative to the impact of weaker competition on auction outcomes. This channel may become more prominent if Kairos attracts additional searchers in the future, particularly those employing more diverse strategies such as atomic arbitrage.

To improve revenue capture, Arbitrum could pursue several directions. Beyond adopting an entirely different mechanism or market structure, the current auction format itself could be modified to better align payments with the value reflected in bids.

First, switching to a first-price auction could make anti-competitive bidding more difficult~\cite{bidder_collusion} and may reduce the gap between value and payments. Second, the auction could respond more effectively to changing demand by adjusting the reserve price dynamically rather than relying on a fixed reserve. In particular, incorporating external price volatility---which appears to be a key driver of searcher activity and auction values---into the reserve-price design could push bids closer to bidders' willingness to pay.

\begin{figure}[t!]
    \centering
    \includegraphics[width=\linewidth]{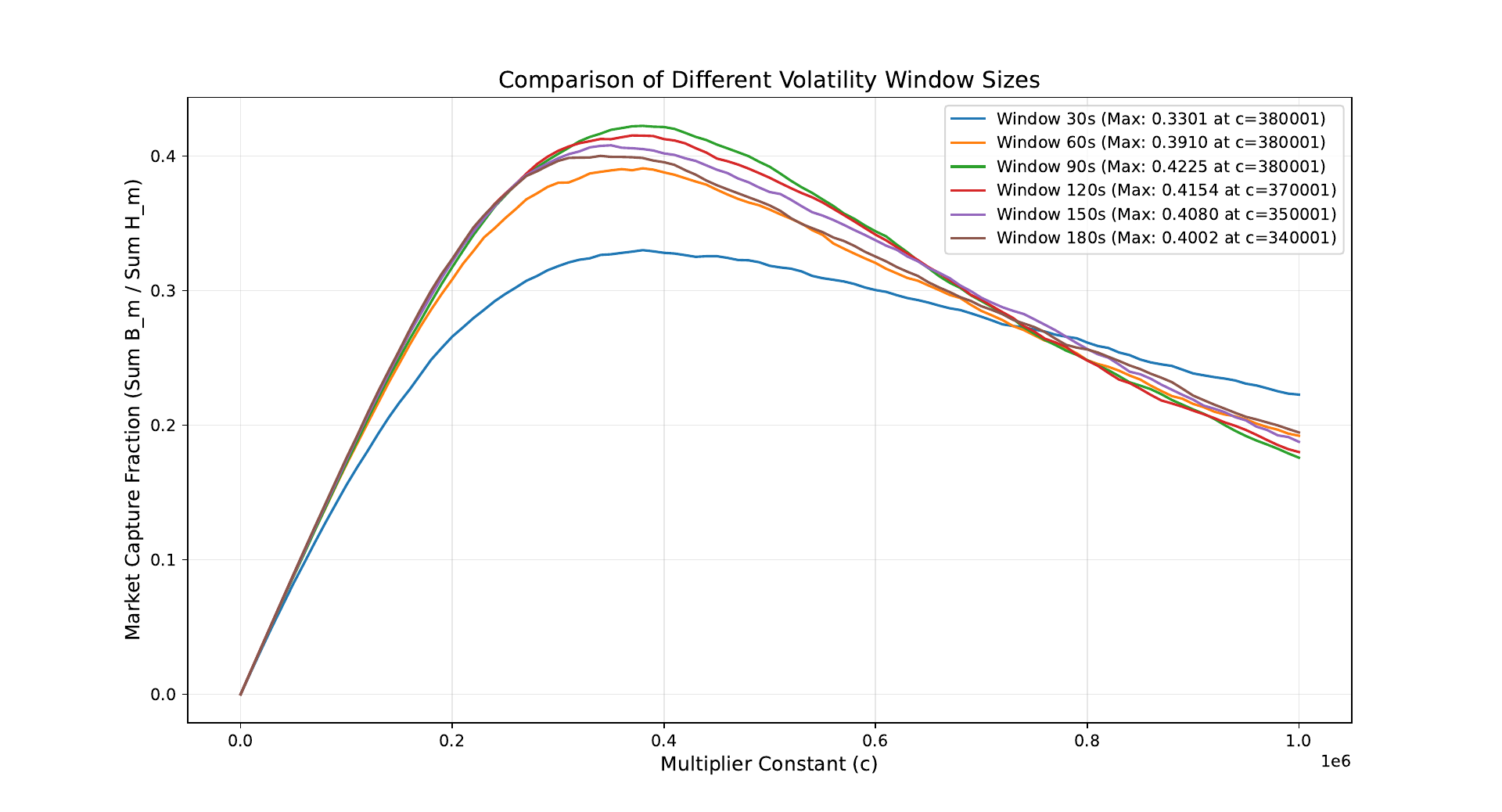}
    \caption{Performance of the dynamic reserve-price rule for different choices of $t$ and $c$, measured relative to the sum of top bids.}
    \label{fig:dynamic_reserve_price}
\end{figure}

We illustrate this with a simple volatility-based reserve-price rule. Specifically, we set the reserve price equal to $c\sigma_t^2$, where $c$ is a constant calibrated using bid data from the \textit{Competition} period and $\sigma_t$ denotes the standard deviation of log returns, $\log(p_{i+1}/p_i)$, computed over a window of length $t$ before the auction ends. Using this rule, we recover at least 80\% of the revenue benchmarked against the sum of top bids during the period from August 2025 to January 2026. \Cref{fig:dynamic_reserve_price} shows the performance of this dynamic reserve-price rule across different choices of $t$ and $c$.

\subsection{Lessons for Ethereum L1}
The lessons from Timeboost's AOT auction may also be relevant for Ethereum L1, subject to two caveats. First, Timeboost is mainly used for CEX--DEX arbitrage, which captures only part of Ethereum MEV, albeit an important and highly latency-sensitive part. A rough analogue on Ethereum would be an AOT auction for top-of-block rights over several consecutive blocks. Second, while revenue capture through Timeboost is a primary concern for Arbitrum, protocol-level MEV capture on Ethereum---for example through MEV-burn-type designs~\cite{drake2023mevburn}---is arguably less central than preserving broader system properties.

With these caveats in mind, two lessons appear particularly relevant. First, on Timeboost the population of dominant searchers is narrow, and the value they extract from the fast-lane is largely concentrated in the same type of opportunity: CEX--DEX arbitrage on a few liquid pairs. On Ethereum L1, MEV opportunities are more heterogeneous and accrue to different searcher classes at different times. An AOT auction allocates ordering rights before this realization is known and is therefore more likely to misallocate them relative to a JIT mechanism. We would expect this allocative inefficiency to be larger on L1, even before any secondary market emerges.

Second, limited competition among dominant searchers can create favorable conditions for secondary markets that move value outside the primary auction. Ethereum's searcher and builder markets remain more competitive than Arbitrum's and are not currently as concentrated as the duopolistic structure we observe here. Still, the searcher market on Ethereum has become more concentrated over time, and similar resale dynamics could plausibly emerge if competition weakens substantially.

More broadly, the Arbitrum case suggests that AOT allocation mechanisms may be particularly vulnerable to secondary-market intermediation once competition among dominant participants becomes weak. This vulnerability may itself be amplified by AOT design. In a JIT auction, valuations are likely to be more dispersed across auctions, allowing participants who are usually not competitive to occasionally outbid dominant players when they identify especially valuable opportunities. Such episodic entry can help preserve some competitive pressure. When ordering rights are sold ahead of time, this source of occasional competitiveness is weaker, potentially making uncompetitive equilibria even easier to sustain.

\bibliography{references}

@article{milionis2022automated,
  title={Automated market making and loss-versus-rebalancing},
  author={Milionis, Jason and Moallemi, Ciamac C and Roughgarden, Tim and Zhang, Anthony Lee},
  journal={arXiv preprint arXiv:2208.06046},
  year={2022}
}

@inproceedings{MEV,
  author       = {Philip Daian and
                  Steven Goldfeder and
                  Tyler Kell and
                  Yunqi Li and
                  Xueyuan Zhao and
                  Iddo Bentov and
                  Lorenz Breidenbach and
                  Ari Juels},
  title        = {Flash Boys 2.0: Frontrunning in Decentralized Exchanges, Miner Extractable
                  Value, and Consensus Instability},
  booktitle    = {2020 {IEEE} Symposium on Security and Privacy, {SP} 2020, San Francisco,
                  CA, USA, May 18-21, 2020},
  pages        = {910--927},
  publisher    = {{IEEE}},
  year         = {2020},
  url          = {https://doi.org/10.1109/SP40000.2020.00040},
  doi          = {10.1109/SP40000.2020.00040},
  bibsource    = {dblp computer science bibliography, https://dblp.org}
}

@InProceedings{mamageishvili_et_al:LIPIcs.AFT.2023.23,
  author =	{Mamageishvili, Akaki and Kelkar, Mahimna and Schlegel, Jan Christoph and Felten, Edward W.},
  title =	{{Buying Time: Latency Racing vs. Bidding for Transaction Ordering}},
  booktitle =	{5th Conference on Advances in Financial Technologies (AFT 2023)},
  pages =	{23:1--23:22},
  series =	{Leibniz International Proceedings in Informatics (LIPIcs)},
  ISBN =	{978-3-95977-303-4},
  ISSN =	{1868-8969},
  year =	{2023},
  volume =	{282},
  editor =	{Bonneau, Joseph and Weinberg, S. Matthew},
  publisher =	{Schloss Dagstuhl -- Leibniz-Zentrum f{\"u}r Informatik},
  address =	{Dagstuhl, Germany},
  URL =		{https://drops.dagstuhl.de/entities/document/10.4230/LIPIcs.AFT.2023.23},
  URN =		{urn:nbn:de:0030-drops-192120},
  doi =		{10.4230/LIPIcs.AFT.2023.23}
}

@inproceedings{schlegel2024transaction,
  title={Transaction Ordering Auctions},
  author={Schlegel, Christoph},
  booktitle={International Conference on Financial Cryptography and Data Security},
  pages={91--104},
  year={2024},
  organization={Springer}
}

@article{mev_capture,
  author       = {Robin Fritsch and
                  Maria In{\^{e}}s Silva and
                  Akaki Mamageishvili and
                  Benjamin Livshits and
                  Edward W. Felten},
  title        = {{MEV} Capture Through Time-Advantaged Arbitrage},
  journal      = {CoRR},
  volume       = {abs/2410.10797},
  year         = {2024},
  url          = {https://doi.org/10.48550/arXiv.2410.10797},
  doi          = {10.48550/ARXIV.2410.10797},
  eprinttype    = {arXiv},
  eprint       = {2410.10797},
  bibsource    = {dblp computer science bibliography, https://dblp.org}
}

@inproceedings{timing_games,
  author       = {Caspar Schwarz{-}Schilling and
                  Fahad Saleh and
                  Thomas Thiery and
                  Jennifer Pan and
                  Nihar Shah and
                  Barnab{\'{e}} Monnot},
  editor       = {Joseph Bonneau and
                  S. Matthew Weinberg},
  title        = {Time Is Money: Strategic Timing Games in Proof-Of-Stake Protocols},
  booktitle    = {5th Conference on Advances in Financial Technologies, {AFT} 2023,
                  October 23-25, 2023, Princeton, NJ, {USA}},
  series       = {LIPIcs},
  volume       = {282},
  pages        = {30:1--30:17},
  publisher    = {Schloss Dagstuhl - Leibniz-Zentrum f{\"{u}}r Informatik},
  year         = {2023},
  url          = {https://doi.org/10.4230/LIPIcs.AFT.2023.30},
  doi          = {10.4230/LIPICS.AFT.2023.30},
  bibsource    = {dblp computer science bibliography, https://dblp.org}
}

@article{cex_dex,
  author       = {Fei Wu and
                  Danning Sui and
                  Thomas Thiery and
                  Mallesh Pai},
  title        = {Measuring {CEX-DEX} Extracted Value and Searcher Profitability: The
                  Darkest of the {MEV} Dark Forest},
  journal      = {CoRR},
  volume       = {abs/2507.13023},
  year         = {2025},
  url          = {https://doi.org/10.48550/arXiv.2507.13023},
  doi          = {10.48550/ARXIV.2507.13023},
  eprinttype    = {arXiv},
  eprint       = {2507.13023},
  bibsource    = {dblp computer science bibliography, https://dblp.org}
}

@article{latency_advantage,
  author       = {Ciamac C. Moallemi and Mallesh M. Pai and Dan Robinson},
  title        = {Latency Advantages in Common-Value Auctions},
  journal      = {CoRR},
  volume       = {abs/2504.02077},
  year         = {2025}
}

@article{auction_theory,
  author       = {Paul R. Milgrom and Robert J. Weber},
  title        = {A THEORY OF AUCTIONS AND COMPETITIVE BIDDING},
  journal      = {Econometrica},
  volume       = {50},
  issue        = {5},
  year         = {1982}
}

@article{essias2025express,
    author = {Johnnatan Messias and Christof Ferreira Torres},
    title = {The Express Lane to Spam and Centralization: An Empirical Analysis of Arbitrum's Timeboost},
    journal = {CoRR},
    volume = {abs/2509.22143},
    year = 2025
}

@article{slot_auction,
    author = {Julian Ma},
    title = {{Block vs. Slot Auction PBS}},
    journal = {Paragraph},
    note = {Available at \url{https://paragraph.com/@julianma/block-vs-slot-auction-pbs}},
    year = 2022
}

@misc{pai2024,
      title={Centralization in Attester-Proposer Separation}, 
      author={Mallesh Pai and Max Resnick},
      year={2024},
      eprint={2408.03116},
      archivePrefix={arXiv},
      primaryClass={econ.TH},
      howpublished={\url{https://arxiv.org/abs/2408.03116}}, 
}

@misc{trusted,
author = {Julian Ma},
title={Trusted Advantage in Slot Auction ePBS},
year={2024},
howpublished ={\url{https://ethresear.ch/t/trusted-advantage-in-slot-auction-epbs/20456}}
}

@online{drake2023mevburn,
  author       = {Justin Drake},
  title        = {MEV burn---a simple design},
  year         = {2023},
  howpublished = {\url{https://ethresear.ch/t/mev-burn-a-simple-design/15590}},
}

@misc{kairos_timeboost_docs,
  author       = {{Gattaca}},
  title        = {Kairos Timeboost Documentation},
  year         = {2026},
  howpublished = {\url{https://docs.kairos-timeboost.xyz/}},
  note         = {Accessed: 2026-03-11}
}

@misc{kairos_parallel_lanes,
  author       = {{Gattaca}},
  title        = {Parallel Kairos Lanes},
  year         = {2026},
  howpublished = {\url{https://docs.kairos-timeboost.xyz/parallel-kairos-lanes}},
  note         = {Accessed: 2026-03-12}
}

@misc{arbitrum_timeboost_gentle_intro,
  author       = {{Arbitrum Foundation}},
  title        = {Timeboost: A Gentle Introduction},
  year         = {2026},
  howpublished = {\url{https://docs.arbitrum.io/how-arbitrum-works/timeboost/gentle-introduction}},
  note         = {Accessed: 2026-03-11}
}

@misc{dune_analytics,
  author       = {{Dune Analytics}},
  title        = {Dune: Blockchain Data Analytics Platform},
  year         = {2026},
  howpublished = {\url{https://dune.com/}},
  note         = {Accessed: 2026-03-11}
}

@misc{binance_market_data,
  author       = {{Binance}},
  title        = {Binance Market Data},
  year         = {2026},
  howpublished = {\url{https://data.binance.vision}},
  note         = {Accessed: 2026-03-11}
}

@misc{flashbots_mevboost,
  author       = {{Flashbots}},
  title        = {What is MEV-Boost?},
  year         = {2026},
  howpublished = {\url{https://docs.flashbots.net/flashbots-mev-boost/introduction}},
  note         = {Accessed: 2026-03-18}
}

@misc{oz2025crosschainarbitragefrontiermev,
      title={Cross-Chain Arbitrage: The Next Frontier of MEV in Decentralized Finance}, 
      author={Burak Öz and Christof Ferreira Torres and Christoph Schlegel and Bruno Mazorra and Jonas Gebele and Filip Rezabek and Florian Matthes},
      year={2025},
      eprint={2501.17335},
      archivePrefix={arXiv},
      primaryClass={cs.CR},
      howpublished = {\url{https://arxiv.org/abs/2501.17335}}, 
}

@misc{arbitrum_timeboost_historical_bids,
  author       = {{Arbitrum Foundation}},
  title        = {How to Use Timeboost: How to View Historical Bid Data},
  year         = {2026},
  howpublished = {\url{https://docs.arbitrum.io/how-arbitrum-works/timeboost/how-to-use-timeboost#how-to-view-historical-bid-data}},
  note         = {Accessed: 2026-03-11}
}

@Article{bidder_collusion,
journal={Journal of Economic Theory},
author={Marshall, Robert C. and Marx, Leslie M.},
title={Bidder collusion},
year={2007},
month={March},
pages={374-402},
volume={133},
number={1},
doi={None},
url={https://ideas.repec.org/a/eee/jetheo/v133y2007i1p374-402.html},
}

@article{timeboost_spam,
  author       = {Agostino Capponi and
                  Brian Zhu},
  title        = {Does Timeboost Reduce MEV-Related Spam? Theory and Evidence from Layer-2
                  Transactions},
  journal      = {CoRR},
  volume       = {abs/2512.10094},
  year         = {2025},
  url          = {https://doi.org/10.48550/arXiv.2512.10094},
  doi          = {10.48550/ARXIV.2512.10094},
  eprinttype   = {arXiv},
  eprint       = {2512.10094},
  bibsource    = {dblp computer science bibliography, https://dblp.org}
}

@article{volatility_mean_reverting,
author = {Boyi Li and Weixuan Xia},
title = {Crypto inverse-power options and fractional stochastic volatility},
journal = {Quantitative Finance},
volume = {25},
number = {7},
pages = {1073--1099},
year = {2025},
publisher = {Routledge},
doi = {10.1080/14697688.2025.2518252},
URL = { 
    https://doi.org/10.1080/14697688.2025.2518252
},
}

@misc{hasbrouck_low-latency_2013,
	location = {Rochester, {NY}},
	title = {Low-Latency Trading},
	url = {https://papers.ssrn.com/abstract=1695460},
	doi = {10.2139/ssrn.1695460},
	number = {1695460},
	publisher = {Social Science Research Network},
	author = {Hasbrouck, Joel and Saar, Gideon},
	urldate = {2026-08-05},
	date = {2013-05-22},
	langid = {english},
}

\appendix
\section{Additional Tables}\label{sec:appendix}




\begin{table}[h]
\centering
\caption{Net markout PnL across alternative horizons for time-boosted trades (August 2025).}
\label{tab:markout_robustness_final}
\begin{tabular}{lrrrrr}
\toprule
& \multicolumn{5}{c}{Markout horizon} \\
\cmidrule(lr){2-6}
Searcher & 1s & 2s & 5s & 10s & 30s \\
\midrule
Wintermute
& 464,792
& 514,990
& 546,884
& 556,499
& 576,517 \\

Selini
& 450,796
& 468,910
& 478,638
& 481,759
& 499,908 \\

\midrule
\textbf{Total}
& \textbf{915,588}
& \textbf{983,900}
& \textbf{1,025,522}
& \textbf{1,038,257}
& \textbf{1,076,425} \\
\bottomrule
\end{tabular}

\vspace{0.4em}
\begin{minipage}{0.92\linewidth}
\footnotesize
\textit{Notes:} The table reports net markout PnL, in USD, for all identified
time-boosted trades in August 2025. The sample is held fixed across horizons
and contains 2,180,087 Wintermute trades and 659,833 Selini trades, for a total
of 2,839,920 observations.
\end{minipage}
\end{table}

\begin{table}[h]
\centering
\caption{Distinct bidding addresses, auction participation, and active periods during the February 1--March 12, 2026 transition-analysis window.}
\label{tab:unique_bidders}

\small
\setlength{\tabcolsep}{5pt}
\begin{tabular}{@{}llrrc@{}}
\toprule
Address & Entity & Bids (share) & Wins (share) & Active$^\dagger$ \\
\midrule
\href{https://arbiscan.io/address/0x95c0d21482fd6bc204e588c06632fdb1cf51b018}
     {\texttt{0x95c0...b018}}
& Selini
& 45,348 (38.71\%)
& 11,268 (23.63\%)
& \texttt{YYYY} \\

\href{https://arbiscan.io/address/0x8c6ffc5501aca218f5f60e140d2c2461092ec3e2}
     {\texttt{0x8c6f...c3e2}}
& Wintermute
& 38,273 (32.67\%)
& 10,302 (21.60\%)
& \texttt{YYYY} \\

\href{https://arbiscan.io/address/0x2b38a73dd32a2eafe849825a4b515ae5187eda42}
     {\texttt{0x2b38...da42}}
& Kairos
& 29,327 (25.03\%)
& 25,968 (54.45\%)
& \texttt{YYYY} \\

\href{https://arbiscan.io/address/0x3141592601bf441700f150bef82db6040f4fe373}
     {\texttt{0x3141...e373}}
& Unknown
& 1,292 (1.10\%)
& 15 (0.03\%)
& \texttt{-{}Y-{}Y} \\

\href{https://arbiscan.io/address/0x4a0acac60321d89e8d4d01fa09318849cb6a586a}
     {\texttt{0x4a0a...586a}}
& Unknown
& 972 (0.83\%)
& 83 (0.17\%)
& \texttt{-{}-{}-{}Y} \\

\href{https://arbiscan.io/address/0x3266692a1b5261e7051766c71b2cca1eeead7599}
     {\texttt{0x3266...7599}}
& Unknown
& 864 (0.74\%)
& 0 (0.00\%)
& \texttt{-{}-{}-{}Y} \\

\href{https://arbiscan.io/address/0x582b91644a8dd808e84cac7c6118a12fcd7cf198}
     {\texttt{0x582b...f198}}
& Unknown
& 406 (0.35\%)
& 6 (0.01\%)
& \texttt{Y-{}-{}-} \\

\href{https://arbiscan.io/address/0x44807b96e244b3f88ea22eed39c4aa28ac255722}
     {\texttt{0x4480...5722}}
& Unknown
& 332 (0.28\%)
& 0 (0.00\%)
& \texttt{Y-{}-{}-} \\

\href{https://arbiscan.io/address/0x2f0707ead77b117384ddd126317b7294767b79aa}
     {\texttt{0x2f07...79aa}}
& Unknown
& 220 (0.19\%)
& 24 (0.05\%)
& \texttt{Y-{}-{}-} \\

\href{https://arbiscan.io/address/0xed217ad2811ca1513381e6f464130803de3f3157}
     {\texttt{0xed21...3157}}
& Unknown
& 95 (0.08\%)
& 19 (0.04\%)
& \texttt{-{}-{}-{}Y} \\

\href{https://arbiscan.io/address/0xda8795ccf620b8559c0ffb695997a3126b8f3b15}
     {\texttt{0xda87...3b15}}
& Unknown
& 6 (0.01\%)
& 0 (0.00\%)
& \texttt{-{}Y-{}-} \\

\href{https://arbiscan.io/address/0x0e1730aab680245971603f9edeaa0c85ebeaaaaa}
     {\texttt{0x0e17...aaaa}}
& Unknown
& 5 (0.00\%)
& 0 (0.00\%)
& \texttt{-{}Y-{}-} \\

\href{https://arbiscan.io/address/0x30c5876c741c4902f638750ee72279b1698b5284}
     {\texttt{0x30c5...5284}}
& Unknown
& 4 (0.00\%)
& 2 (0.00\%)
& \texttt{-{}-{}-{}Y} \\

\href{https://arbiscan.io/address/0xb79e316083981fe186f60f81d8a20010ba96ca6c}
     {\texttt{0xb79e...ca6c}}
& Unknown
& 2 (0.00\%)
& 2 (0.00\%)
& \texttt{-{}Y-{}-} \\

\midrule
\textbf{Total}
& {}
& \textbf{117,146 (100.0\%)}
& \textbf{47,689 (100.0\%)}
& {} \\
\bottomrule
\end{tabular}

\vspace{3pt}
\begin{minipage}{0.98\linewidth}
\footnotesize
\raggedright
\textit{Notes:} $^\dagger$Activity flags correspond, in order, to the Competition, Kairos,
Reserve Price Adaptation, and Steady State periods.
\texttt{Y} indicates at least one submitted bid; \texttt{-} indicates no bids.
\end{minipage}
\end{table}

\begin{table}[h]
\centering
\small
\caption{Pearson correlations between ETH hourly volatility and hourly Timeboost auction metrics during the February 1--March 12, 2026 transition-analysis window. Metrics are denominated in ETH and reported by transition sub-period. $n$ denotes the number of hours in which at least one bid was submitted. Significance: $^{*}p<0.05$, $^{**}p<0.01$, $^{***}p<0.001$.}
\label{tab:vol_corr_eth}
\begin{tabular}{llrrr}
\toprule
Metric & Zone & $r$ & $p$ & $n$ \\
\midrule
\multirow{4}{*}{Paid Bid (ETH)}
  & Competition         & $+0.805$ & $0.000^{***}$ & 285 \\
  & Kairos              & $+0.225$ & $0.007^{**}$  & 145 \\
  & Res.\ Price Adapt.  & $+0.686$ & $0.000^{***}$ &  24 \\
  & Steady State        & $+0.828$ & $0.000^{***}$ & 365 \\
\midrule
\multirow{4}{*}{Top Bid (ETH)}
  & Competition         & $+0.877$ & $0.000^{***}$ & 285 \\
  & Kairos              & $-0.036$ & $0.665$       & 145 \\
  & Res.\ Price Adapt.  & $+0.482$ & $0.017^{*}$   &  24 \\
  & Steady State        & $+0.837$ & $0.000^{***}$ & 365 \\
\midrule
\multirow{4}{*}{Absolute Bid Gap (ETH)}
  & Competition         & $+0.773$ & $0.000^{***}$ & 285 \\
  & Kairos              & $-0.233$ & $0.005^{**}$  & 145 \\
  & Res.\ Price Adapt.  & $+0.382$ & $0.065$       &  24 \\
  & Steady State        & $+0.838$ & $0.000^{***}$ & 365 \\
\bottomrule
\end{tabular}
\end{table}

\begin{table}[htbp]
  \centering
  \small
\caption{Distribution of bid amounts during the February 1--March 12, 2026 transition-analysis window, by bidder and transition sub-period. Bid amounts are denominated in ETH. $n$ denotes the number of bids submitted.}
  \label{tab:bid_distribution}
  \begin{tabular}{ll rrrrrr}
    \toprule
    Zone & Bidder & $n$ & p25 & Median & p75 & p99 & Mean \\
    \midrule
    \textit{Competition} & Wintermute & 15,625 & 0.0034 & 0.0075 & 0.0162 & 0.0871 & 0.0136 \\
                        & Selini     & 16,863 & 0.0037 & 0.0062 & 0.0127 & 0.0602 & 0.0108 \\
                        & Kairos     &    307 & 0.0044 & 0.0069 & 0.0095 & 0.0168 & 0.0074 \\
    \midrule
    \textit{Steady State} & Wintermute & 21,537 & 0.0012 & 0.0027 & 0.0032 & 0.0096 & 0.0025 \\
                          & Selini     & 20,043 & 0.0015 & 0.0021 & 0.0033 & 0.0102 & 0.0029 \\
                          & Kairos     & 20,382 & 0.0087 & 0.0182 & 0.0234 & 0.0703 & 0.0198 \\
    \bottomrule
  \end{tabular}
\end{table}

\begin{table}[h]
\centering
\small
\caption{Estimated positive-PnL CEX--DEX arbitrage during the February 1--March 12, 2026 transition-analysis window, by searcher, transition sub-period, and time-boost status. The table reports transaction counts, total PnL, and average PnL per transaction for time-boosted and regular transactions.}
\label{tab:pnl_summary_zone}
\resizebox{\textwidth}{!}{%
\begin{tabular}{llrrrrrr}
\toprule
 & & \multicolumn{3}{c}{Time-boosted} & \multicolumn{3}{c}{Regular} \\
\cmidrule(lr){3-5}\cmidrule(lr){6-8}
Searcher & Zone & TXs & Total PnL (\$) & Avg (\$) & TXs & Total PnL (\$) & Avg (\$) \\
\midrule
\multirow{5}{*}{Wintermute}
 & Competition         & 461{,}590 & 2{,}098{,}400 & 4.55 & 300{,}987 & 1{,}127{,}281 & 3.75 \\
 & Kairos              & 228{,}081 &   415{,}139   & 1.82 &   2{,}090 &     2{,}848   & 1.36 \\
 & Res.\ Price Adapt.  &  21{,}353 &    37{,}759   & 1.77 & 255{,}053 &   471{,}415   & 1.85 \\
 & Steady State        & 678{,}616 & 1{,}370{,}977 & 2.02 &  87{,}054 &   177{,}372   & 2.04 \\
\cmidrule{2-8}
 & \textit{Total}      & \textit{1{,}389{,}640} & \textit{3{,}922{,}275} & \textit{2.82} & \textit{645{,}184} & \textit{1{,}778{,}917} & \textit{2.76} \\
\midrule
\multirow{5}{*}{Selini}
 & Competition         &  79{,}875 &   484{,}942   & 6.07 &  99{,}536 &   672{,}907   & 6.76 \\
 & Kairos              &  39{,}723 &   168{,}443   & 4.24 &   3{,}597 &    17{,}400   & 4.84 \\
 & Res.\ Price Adapt.  &   1{,}152 &     4{,}685   & 4.07 &  41{,}011 &   196{,}012   & 4.78 \\
 & Steady State        & 162{,}495 &   890{,}996   & 5.48 &  18{,}114 &    92{,}114   & 5.09 \\
\cmidrule{2-8}
 & \textit{Total}      & \textit{283{,}245} & \textit{1{,}549{,}067} & \textit{5.47} & \textit{162{,}258} & \textit{978{,}434} & \textit{6.03} \\
\bottomrule
\end{tabular}}
\end{table}

\begin{table}[h]
\centering
\small
\caption{Pearson correlations between ETH hourly volatility and hourly searcher PnL during the February 1--March 12, 2026 transition-analysis window, by PnL metric and transition sub-period. Results are restricted to positive-PnL CEX--DEX arbitrage transactions. $n$ denotes the number of hours with at least one transaction. TB = time-boosted; Non-TB = regular, non-time-boosted. Significance: $^{*}p<0.05$, $^{**}p<0.01$, $^{***}p<0.001$.}
\label{tab:pnl_vol_corr}
\begin{tabular}{lllrrr}
\toprule
Searcher & Metric & Zone & $r$ & $p$ & $n$ \\
\midrule
\multirow{12}{*}{Wintermute}
 & \multirow{4}{*}{TB PnL}
   & Competition         & $+0.870$ & $0.000^{***}$ & 285 \\
 & & Kairos              & $+0.897$ & $0.000^{***}$ & 144 \\
 & & Res.\ Price Adapt.  & $+0.454$ & $0.000^{***}$ & 167 \\
 & & Steady State        & $+0.791$ & $0.000^{***}$ & 363 \\
\cmidrule{2-6}
 & \multirow{4}{*}{Non-TB PnL}
   & Competition         & $+0.843$ & $0.000^{***}$ & 285 \\
 & & Kairos              & $+0.183$ & $0.028^{*}$   & 144 \\
 & & Res.\ Price Adapt.  & $+0.785$ & $0.000^{***}$ & 167 \\
 & & Steady State        & $+0.704$ & $0.000^{***}$ & 363 \\
\cmidrule{2-6}
 & \multirow{4}{*}{Total PnL}
   & Competition         & $+0.917$ & $0.000^{***}$ & 285 \\
 & & Kairos              & $+0.902$ & $0.000^{***}$ & 144 \\
 & & Res.\ Price Adapt.  & $+0.857$ & $0.000^{***}$ & 167 \\
 & & Steady State        & $+0.822$ & $0.000^{***}$ & 363 \\
\midrule
\multirow{12}{*}{Selini}
 & \multirow{4}{*}{TB PnL}
   & Competition         & $+0.788$ & $0.000^{***}$ & 284 \\
 & & Kairos              & $+0.818$ & $0.000^{***}$ & 144 \\
 & & Res.\ Price Adapt.  & $+0.294$ & $0.000^{***}$ & 167 \\
 & & Steady State        & $+0.868$ & $0.000^{***}$ & 359 \\
\cmidrule{2-6}
 & \multirow{4}{*}{Non-TB PnL}
   & Competition         & $+0.679$ & $0.000^{***}$ & 284 \\
 & & Kairos              & $+0.179$ & $0.032^{*}$   & 144 \\
 & & Res.\ Price Adapt.  & $+0.804$ & $0.000^{***}$ & 167 \\
 & & Steady State        & $+0.473$ & $0.000^{***}$ & 359 \\
\cmidrule{2-6}
 & \multirow{4}{*}{Total PnL}
   & Competition         & $+0.712$ & $0.000^{***}$ & 284 \\
 & & Kairos              & $+0.830$ & $0.000^{***}$ & 144 \\
 & & Res.\ Price Adapt.  & $+0.816$ & $0.000^{***}$ & 167 \\
 & & Steady State        & $+0.872$ & $0.000^{***}$ & 359 \\
\bottomrule
\end{tabular}
\end{table}

\end{document}